\documentclass[11pt]{article}
\usepackage{epsfig}
\usepackage{amsfonts}
\usepackage{amsmath}
\usepackage{bbm,bm}
\usepackage{cite}
\allowdisplaybreaks[1]

\usepackage{tikz}
\usetikzlibrary{shapes,arrows,shadows,automata,positioning}
\newdimen\nodeDist
\usepackage[
  colorlinks=true,
  linkcolor=black,
  citecolor=black,
  filecolor=black,
  urlcolor=black,
  breaklinks=true
  ]{hyperref}

\newcommand{\nocontentsline}[3]{}
\newcommand{\tocless}[2]{\bgroup\let\addcontentsline=\nocontentsline#1{#2}\egroup}

\newcommand{\MSbar}{$\overline{\text{MS}}$\;}
\newcommand{\Lamd}{\mu_{3}}
\newcommand{\LamD}{\mu}

\newcommand{\nn}{\nonumber \\}
\newcommand{\rmi}[1]{{\mbox{\scriptsize #1}}}
\newcommand{\rmii}[1]{{\mbox{\tiny\rm{#1}}}}

\newcommand{\MW}{M_\rmii{$W$}}
\newcommand{\MZ}{M_\rmii{$Z$}}
\newcommand{\mW}{m_\rmii{$W$}}
\newcommand{\mZ}{m_\rmii{$Z$}}
\newcommand{\mA}{m_\rmii{$A$}}

\newcommand{\gY}{g_{\rmii{$Y$}}}
\newcommand{\gs}{g_\rmi{s}}

\newcommand{\GF}{G_\rmi{F}}

\newcommand{\Tc}{T_{\rm c}}

\makeatletter \@addtoreset{equation}{section} \makeatother
\renewcommand{\theequation}{\arabic{section}.\arabic{equation}}
\makeatletter
\renewcommand\section{\@startsection{section}{1}{\z@}%
  {-5.5ex \@plus -1ex \@minus -.2ex}
  {2.3ex \@plus.2ex}%
  {\normalfont\large\bfseries}}
\renewcommand\subsection{\@startsection{subsection}{2}{\z@}%
  {-3.25ex\@plus -1ex \@minus -.2ex}%
  {1.5ex \@plus .2ex}%
  {\normalfont\normalsize\bfseries}}
\renewcommand\thesection{\@arabic\c@section}
\renewcommand\thesubsection{\thesection.\@arabic\c@subsection}
\renewcommand{\@seccntformat}[1]{%
  \csname the#1\endcsname.\hspace{1.0em}}
\makeatother

\begin{document}

\flushbottom

\begin{titlepage}

\begin{flushright}
HIP-2022-2/TH\\
NORDITA 2022-009\\
\end{flushright}
\begin{centering}

\vfill

{\Large{\bf
Combining thermal resummation and gauge invariance for electroweak phase transition
}}

\vspace{0.8cm}

\renewcommand{\thefootnote}{\fnsymbol{footnote}}
Philipp Schicho$^{\rm a,}$%
\footnote{philipp.schicho@helsinki.fi},
Tuomas V.~I.~Tenkanen$^{\rm b,c,d,}$%
\footnote{tuomas.tenkanen@su.se}
and
Graham White$^{\rm e,}$%
\footnote{graham.white@ipmu.jp}

\vspace{0.8cm}

$^{\rm a}$%
{\em
Department of Physics and Helsinki Institute of Physics,\\
P.O.\ Box 64,
FI-00014 University of Helsinki,
Finland\\}
\vspace{0.3cm}

$^\rmi{b}$%
{\em
Nordita,
KTH Royal Institute of Technology and Stockholm University,\\
Roslagstullsbacken 23,
SE-106 91 Stockholm,
Sweden\\}
\vspace{0.3cm}

$^\rmi{c}$%
{\em
Tsung-Dao Lee Institute \& School of Physics and Astronomy,\\
Shanghai Jiao Tong University, Shanghai 200240, China\\}
\vspace{0.3cm}

$^\rmi{d}$%
{\em
Shanghai Key Laboratory for Particle Physics and Cosmology, Key Laboratory for Particle Astrophysics and Cosmology (MOE), Shanghai Jiao Tong University,\\
Shanghai 200240, China\\}
\vspace{0.3cm}

$^\rmi{e}$%
{\em
Kavli IPMU (WPI), UTIAS, The University of Tokyo, Kashiwa, Chiba 277-8583, Japan\\}

\vspace*{0.8cm}

\mbox{\bf Abstract}

\end{centering}

\vspace*{0.3cm}

\noindent

For computing thermodynamics of
the electroweak phase transition,
we discuss a minimal approach that reconciles both
gauge invariance and
thermal resummation. 
Such a minimal setup
consists of
a two-loop dimensional reduction to three-dimensional effective theory,
a one-loop computation of the effective potential and
its expansion around the leading-order minima within the effective theory. 
This approach is tractable and provides formulae for resummation that are arguably no more complicated than those that appear in standard 
techniques ubiquitous in the literature.
In particular, we implement renormalisation group improvement related to the hard thermal scale.
Despite its generic nature, we present this approach for
the complex singlet extension of the Standard Model
which has interesting prospects for high energy collider phenomenology 
and
dark matter predictions.
The presented expressions can be used in future studies of
phase transition thermodynamics and
gravitational wave production. 

\vfill
\end{titlepage}

\tableofcontents
\clearpage

\renewcommand{\thefootnote}{\arabic{footnote}}
\setcounter{footnote}{0}

%
\section{Introduction}
\label{sec:intro}

Understanding the thermal history
of electroweak symmetry breaking is
one of the central endeavours of next generation experiments, with both
next generation colliders~\cite{%
  Ramsey-Musolf:2019lsf,Papaefstathiou:2020iag,Chala:2018opy} 
and 
gravitational wave detectors~\cite{Caprini:2019egz} 
potentially giving definitive answers. 
Furthermore, if electroweak symmetry breaking occurs via 
a strong first-order phase transition, it could answer one of
the central issues of cosmology -- why is there more matter than anti-matter in
the present day universe~\cite{Morrissey:2012db,White:2016nbo}.
Since the Standard Model (SM) of particle physics predicts
a smooth crossover transition~\cite{%
  Kajantie:1995kf,Kajantie:1996mn,Kajantie:1996qd,Csikor:1998eu,DOnofrio:2015gop},
modifying this expectation necessitates new 
beyond the Standard Model (BSM) states at the electroweak scale~\cite{%
  Pietroni:1992in,Cline:1996mga,Ham:2004nv,Funakubo:2005pu,Barger:2008jx,Chung:2010cd,
  Espinosa:2011ax,Chowdhury:2011ga,Gil:2012ya,Carena:2012np,No:2013wsa,Dorsch:2013wja,
  Curtin:2014jma,Huang:2014ifa,Profumo:2014opa,Kozaczuk:2014kva,Jiang:2015cwa,
  Curtin:2016urg,Vaskonen:2016yiu,Dorsch:2016nrg,Huang:2016cjm,Chala:2016ykx,
  Basler:2016obg,Beniwal:2017eik,Bernon:2017jgv,Kurup:2017dzf,Andersen:2017ika,
  Chiang:2017nmu,Dorsch:2017nza,Beniwal:2018hyi,Bruggisser:2018mrt,Athron:2019teq,
  Kainulainen:2019kyp,Bian:2019kmg,Li:2019tfd,Chiang:2019oms,Xie:2020bkl,Bell:2020gug}.
In the presence of a global symmetry, among such new states could in fact be
a dark matter candidate~\cite{%
  Burgess:2000yq,Gonderinger:2012rd,Chao:2014ina,Jiang:2015cwa,Chiang:2017nmu},
giving a minimal unified explanation for the origin of matter.

Describing the electroweak phase transition (EWPT) perturbatively is
a theoretical challenge.
In non-Abelian gauge theories at finite temperature, perturbation theory fails completely 
at high enough orders~\cite{Linde:1980ts}. 
Even at lower orders, the effective expansion parameter can be
large despite
a weakly coupled zero-temperature theory.
Physically, this is a consequence of enhancement of light bosonic modes at
the infrared (IR), and for a perturbative description to converge requires thermal resummations.
The 
thermal plasma exhibits a class of mass hierarchies at high temperature, and in 
perturbation theory this can be described by integrating out
contributions of the heavy thermal scale to parameters of
an effective theory at lower scales~\cite{Arnold:1992fb}.
This idea of effective descriptions 
is systematised by
high-temperature dimensional reduction~\cite{Ginsparg:1980ef, Appelquist:1981vg}
to three-dimensional effective field theory (3d EFT)~\cite{Braaten:1995cm,Kajantie:1995dw}. 
Dimensional reduction allows to by-pass problems of perturbation theory at the infrared 
using non-perturbative lattice simulations of
the 3d EFT~\cite{Farakos:1994xh,Kajantie:1995kf}.  
Due to the excessive cost and technical effort of such simulations,
the virtue of perturbation theory within the 3d EFT~\cite{Farakos:1994kx} 
is to guide thermodynamic investigations. 

Past decade
EWPT analyses most often
focused on minimising the thermal effective potential
that adopts the resummation of
``the most dangerous daisy diagrams''.
At leading order,
this is achieved by
resumming one-loop thermal corrections to masses of zero Matsubara modes and
subsequently computing the one-loop effective potential for the resummed zero modes.
The 3d EFT picture translates this to
a dimensional reduction at leading order
(one-loop in effective masses, 
tree-level in couplings) and
a computation of the one-loop effective potential within the EFT.
On the other hand, 
this description fails to include several important
two-loop contributions~\cite{Arnold:1992fb,Arnold:1992rz}.
Without these
two-loop contributions, leading renormalisation group improvement is 
incomplete~\cite{Arnold:1992fb,Farakos:1994kx,Gould:2021oba}.
Such an improvement 
cancels the renormalisation scale between the running of leading-order 
contributions and explicit logarithmic terms at next-to-leading order (NLO).
Indeed, due to the slower convergence of perturbation theory at high temperature,
this cancellation requires a two-loop level computation as
a subset of leading logarithms only appears at this loop level. 
Recent work~\cite{Kainulainen:2019kyp,Croon:2020cgk,Niemi:2021qvp,Gould:2021oba}
has stressed the numerical importance of a consistent perturbative computation to
account for these effects in BSM scenarios of the electroweak phase transition. 

Conventional 
methods for studying the electroweak phase transition in BSM theories suffer from an undesired gauge dependence~\cite{Patel:2011th}
during a direct numerical minimisation of the daisy-resummed thermal effective potential.
One established method for removing this gauge dependence is to perform an 
``$\hbar$-expansion''.
There,
the effective potential is expanded around its leading-order minima~\cite{Laine:1994zq,Patel:2011th}.
In this expansion, the Nielsen identities~\cite{Nielsen:1975fs,Fukuda:1975di}
guarantee gauge independence.  

In the PRM scheme~\cite{Patel:2011th},
gauge invariance comes at the cost of neglecting 
important resummations of infrared enhanced diagrams when 
calculated at 
$\mathcal{O}(\hbar)$.
The article at hand resolves this technical issue and 
describes an improved approach, where
(A) thermal resummations are handled by the dimensionally reduced 3d EFT with
(B) an $\hbar$-expansion applied within the EFT. 
Gauge invariance is ensured at each step individually.
This approach was originally devised
in~\cite{Laine:1994zq} and
recently applied in~\cite{Niemi:2020hto,Croon:2020cgk,Gould:2021oba}
(also cf.~\cite{Laine:2015kra}). 
However, here we present important updates: 
for radiatively generated transitions,
we resum
one-loop contributions from heavy fields together with
tree-level terms to the leading order potential~\cite{%
  Arnold:1992rz,Ekstedt:2020abj,Gould:2021ccf,Hirvonen:2021zej}.
Hence, the barrier required for a first-order phase transition is already present 
at leading order (LO) in perturbation theory.
This can furthermore help to
avoid a spurious infrared divergence encountered at
higher orders~\cite{Laine:1994zq,Niemi:2020hto}.
In particular, we advocate a minimal%
\footnote{
  An even more bare bones setup is dimensional reduction
  at leading order 
  (one-loop in masses, 
  tree-level in couplings).
  However, the resulting EFT cannot provide RG improvement for the hard thermal scale.
}
setup with NLO
dimensional reduction to 3d EFT 
(two-loop in masses, 
one-loop in couplings), and
a one-loop computation of the effective potential within the EFT.
When determining
the critical temperature ($\Tc$) and
other thermodynamic quantities of interest,
we resort to a technique described in~\cite{Patel:2011th,Niemi:2020hto} rather than 
expanding $\Tc$ explicitly in $\hbar$ as in~\cite{Laine:1994zq}.

All ingredients of our approach have appeared individually in
earlier literature~\cite{%
  Farakos:1994kx,Laine:1994zq,Kajantie:1995dw,Braaten:1995cm,Patel:2011th,Ekstedt:2020abj,Gould:2021ccf,Hirvonen:2021zej}.
Here, we 
uniquely apply them together for the first time
with a BSM application.
An another similar combination of
thermal resummation and
gauge invariance was recently developed~\cite{Ekstedt:2020abj}.
This reference
uses
a direct computation with consistent power counting to ensure gauge invariance
instead of the 3d EFT approach.
The power of the 3d EFT approach has been further demonstrated
in a novel technique for thermal bubble nucleation~\cite{Gould:2021ccf}
which was applied in~\cite{Hirvonen:2021zej}
(cf.~\cite{Ekstedt:2021kyx,Ekstedt:2022tqk} and related~\cite{Garny:2012cg,Lofgren:2021ogg}).
Principally, the approach suggested in the article at hand can be 
embedded in a more generic framework~\cite{Gould:2021ccf}.

The proposed approach applies to a wide range of BSM theories of
the electroweak phase transition. 
One such theory is
the complex-singlet extension of the Standard Model (cxSM)~\cite{Barger:2008jx}
for which we present our prescription.
The model contains two new scalar states,
one that could form a stable dark matter candidate and
one that participates dynamically in the phase transition.
This renders
the model an appealing minimal BSM candidate for both
dark matter and
a potentially strong electroweak phase transition.   
Recent literature has debated of how to organise
the perturbative computation of cxSM thermodynamics.
Ref.~\cite{Jiang:2015cwa}
includes leading daisy resummations,%
\footnote{
  In particular,
  this reference uses Parwani resummation~\cite{Parwani:1991gq} which resumms
  the masses of all bosonic modes and not just those for
  the zero Matsubara modes as in~\cite{Arnold:1992rz,Kajantie:1995dw,Braaten:1995cm}. 
}
but
lacks gauge invariance, while
ref.~\cite{Chiang:2017nmu}
maintains gauge invariance at
the cost of resummation (also cf.~\cite{Chen:2019ebq} and
a more recent investigation~\cite{Cho:2021itv}). 
Similar confusion can in principle arise in any BSM model.
Our computation combines
the benefits of both previous studies and
also improves them by removing their aforementioned shortcomings. 
Consistent with~\cite{Croon:2020cgk,Gould:2021oba},
we find significant error arising from omitting two-loop corrections to thermal masses.
The takeaway message of this article is that
such corrections should not be omitted.
We furthermore demonstrate how to implement them in a gauge-invariant manner.

The structure of this article is as follows.
Section~\ref{sec:setup} 
summarises our approach,
introduces the model and its 3d EFT at high temperature, and
describes the computation of thermodynamics.
In addition, we review the perturbative expansion for the effective potential.
In section~\ref{sec:numerics} we numerically demonstrate our approach at
a representative benchmark point.
Section~\ref{sec:discussion} summarises our findings and discusses their application
for potential future parameter space scans.
Appendix~\ref{sec:details}
collects technical details and
explicates formulae of our analysis.
Finally,
appendix~\ref{sec:light-higgs} discusses renormalisation group improvement
and an epilogue therein exemplifies
the importance of including thermal masses at two-loop order,
in a toy model of the SM with an artificially light Higgs.

%
\section{Thermodynamics: resummation and gauge invariance}
\label{sec:setup}

\begin{figure}
\centering
\tikzstyle{zeroT}=[fill=white!20,minimum width=3.0em,
    align=center,minimum height=2.5em]
\def\blockw{3}
\def\blockh{1.5}
\newcommand\circled[1]{%
  \tikz[baseline=(X.base)]
    \node (X) [shape=circle,inner sep=-1pt,fill=white,text=black] {\strut\scriptsize #1};%
  }
\begin{tikzpicture}
  \node (lag4d) [zeroT]  {$\mathcal{L}_{\rmii{4d}}$};
  \path (lag4d.east)+(\blockw,0) node (lag3d) [zeroT] {$\mathcal{L}_{\rmii{3d}}$};
  \path (lag3d.east)+(\blockw,0) node (veff3d) [zeroT] {$V^{\rmii{3d}}_{\rmii{eff}}$};
  \path (veff3d.east)+(\blockw,0) node (thermo) [zeroT]
    {Thermodynamics \\$\bigl\{\Tc^{ },L/\Tc^4,\dots\bigr\}$};
  \draw[->,>=latex] (lag4d.east) -- node[midway]
    {\hyperref[it:th:1]{\circled{$(1)$}}} (lag3d.west);
  \draw[->,>=latex] (lag3d.east) -- node[midway]
    {\hyperref[it:th:2]{\circled{$(2)$}}} (veff3d.west);
  \draw[->,>=latex] (veff3d.east) -- node[midway]
    {\hyperref[it:th:3]{\circled{$(3)$}}} (thermo.west);
\end{tikzpicture}
\caption{%
  Schematic illustration of a perturbative 3d EFT computation of thermodynamics.
  This article advocates a novel prescription for combining 
  $(1) \to (2) \to (3)$ (cf.\ sec.~\ref{sec:setup}).
  This specific implementation aligns with 
  the generic steps $(d) \to (e) \to (f)$ in fig.~2 in~\cite{Schicho:2021gca}.
}
\label{fig:pipeline-thermo}
\end{figure}

This technical section details our computation.
In fig.~\ref{fig:pipeline-thermo},
we schematically outline the computation of thermodynamics,
adapted from~\cite{Schicho:2021gca}.
The novel prescription suggested in this article combines
\begin{itemize}
  \item[(1)]
    \label{it:th:1}
    Dimensional reduction to 3d EFT at NLO~\cite{Kajantie:1995dw}
    which determines thermal masses up-to two-loo level.
  \item[(2)]
    \label{it:th:2}
    Computing the effective potential within
    the 3d EFT~\cite{Farakos:1994kx} at one-loop level and in
    $\hbar$-expansion~\cite{Laine:1994zq}.
    For a radiative barrier, heavy field contributions at one-loop level are resummed to
    the potential at
    LO~\cite{Gould:2021ccf,Hirvonen:2021zej,Ekstedt:2022yyy,Gould:2022xxx}.
  \item[(3)]
    \label{it:th:3}
    Determining
    critical temperatures numerically from a condition that 
    the value of the effective potential at different phases is
    degenerate~\cite{Patel:2011th} as well as
    determining
    gauge-invariant condensates and latent heat~\cite{Farakos:1994xh,Niemi:2020hto}. 
\end{itemize}

For~(1),
it is crucial that thermal resummation by
dimensional reduction is performed at NLO, instead of LO.
The NLO corrections are sizable, which is already indicated by
a large RG scale dependence of the LO computation~\cite{Croon:2020cgk,Gould:2021oba}. 
Typical computations of the thermal effective potential at one-loop level 
resum also masses at one-loop. 
Such resummation matches the accuracy of the LO dimensional reduction, and is
hence insufficient to eliminate large RG scale dependence. 

In~(2),
we have chosen practicality over full RG improvement.
The computation of the 3d effective potential is straightforward at one-loop,
as one only needs background field dependent mass eigenvalues.
However, RG improvement related to the 3d EFT RG scale can only be eliminated at
two-loop order~\cite{Farakos:1994kx}.
Naturally, such a computation is more challenging
(cf.~\cite{Laine:1994zq,Niemi:2020hto,Croon:2020cgk,Schicho:2021gca}). 
However, we demonstrate that in practice the dependence on
an uneliminated 3d RG scale is less severe as
the aforementioned dependence on 4d RG scale --
a trend demonstrated already in~\cite{Croon:2020cgk}.  

In~(3),
our numerical implementation
follows~\cite{Patel:2011th,Niemi:2020hto} and
differs from the strategy in~\cite{Laine:1994zq,Croon:2020cgk}.
There,
the critical temperature itself is expanded in $\hbar$ and solved order-by-order.

The remainder of this section details our
conventions for the complex-singlet extension of the Standard Model (cxSM),
the structure of the corresponding 3d EFT,
a review of the perturbative expansion,
the thermal effective potential within the EFT and
a gauge-invariant computation of the thermodynamic quantities of interest.
Several formulae are collected in appendix~\ref{sec:details},
for a reader to replicate our analysis.  
While our discussion and concrete expressions focus on 
the cxSM, we present the underlying ideas generically,
for them to be applied in other BSM theories.
In particular, we advocate that our prescription could be implemented in
software for analysing
the electroweak phase transition, such as
{\tt CosmoTransitions}~\cite{Wainwright:2011kj},
{\tt BSMPT}~\cite{Basler:2018cwe,Basler:2020nrq}, and
{\tt PhaseTracer}~\cite{Athron:2020sbe}.   

%
\subsection{A complex-singlet extended Standard Model}

We use the model of ref.~\cite{Chiang:2017nmu} 
with the Standard Model augmented by a complex singlet scalar $S$,
abbreviated as the cxSM.
The scalar sector of the full 4d Lagrangian in Euclidean space reads  
\begin{align}
\label{eq:lag-4d}
\mathcal{L}^{\text{4d}}_\text{scalar} &= 
      (D_\mu\phi)^\dagger (D_\mu\phi)
    + \mu_h^2\phi^\dagger \phi+\lambda_h(\phi^\dagger \phi)^2
    \nn &
    + \frac{1}{2}(\partial_\mu S^*)(\partial_\mu S)
    + \frac{1}{2} b_2 S^*S
    + \frac{1}{4} d_2 (S^*S)^2
    \nn &
    + \frac{1}{2} \delta_2 S^*S \phi^\dagger \phi
    + a_1 S + a^*_1 S^*
    + \frac{1}{4}\Big(b_1 SS + b^*_1 S^*S^* \Big)
    \;.
\end{align}
The scalar fields are parametrised as
\begin{align}
\label{eq:fields}
\phi &=
\begin{pmatrix}
  G^+ \\ 
  \frac{1}{\sqrt{2}} (v_0 + h + i z) 
\end{pmatrix}
\;,
\quad \quad
S = \frac{1}{\sqrt{2}}(v_{S_0} + s + i A )
\;,
\end{align}
where
$v_0$ and $v_{S_0}$ are zero-temperature real vacuum-expectations-values (VEV) and
$G^- = (G^+)^\dagger$.
After electroweak symmetry breaking,
this model has three scalar eigenstates:
two CP-even eigenstates resulting from mixing $h$ and $s$, and
one CP-odd state, $A$. 
All complex phases that can mix $s$ and $A$ are assumed to be zero.
The SM-like Higgs boson is one of the CP-even fields, and
$A$ is the dark matter candidate. 
The fields
$G^{\pm}$ and
$z$ are the usual Nambu-Goldstone bosons as in the SM. 
A tree-level analysis of relations between
the Lagrangian parameters above (in \MSbar-scheme) and
the parameters of the mass eigenstate basis can be found in~\cite{Chiang:2017nmu}
and we merely collect the relevant relations in appendix~\ref{sec:MSbars}.
We follow conventions of~\cite{Brauner:2016fla,Schicho:2021gca} for the SM field content.
The one-loop renormalisation group equations for the \MSbar-parameters are collected
in appendix~\ref{sec:betas}. 
Below, we compute selected thermodynamic properties in terms of \MSbar-parameters of
the Lagrangian.
Only when turning to numerics,
we express
these parameters in terms of physical quantities.

%
\subsection{High temperature 3d EFT}

At sufficiently high temperatures,
the equilibrium thermodynamics of a weakly interacting quantum field theory
(with weak coupling constant, $g$) in the Matsubara formalism can be described by
a dimensionally reduced effective theory (cf.~\cite{Braaten:1995cm,Kajantie:1995dw}). 
Therein the three-dimensional zero Matsubara modes at
the soft/light scale (with masses $\sim g T$) are screened by
the hard/heavy non-zero Matsubara modes (with masses $\sim \pi T$). 
Technically, we shall work at the ultrasoft scale ($g^2 T$) EFT.
This means that in addition to
the hard non-zero Matsubara modes, also
the soft temporal scalar fields have been integrated out%
\footnote{
  The heat bath breaks the Lorentz invariance in the 4d parent theory.
  Hence, temporal components of the gauge fields are represented by
  the temporal scalar sector within the 3d EFT~\cite{Kajantie:1995dw}.
  Since the singlet does not couple to gauge fields directly, but only via Higgs loops, contributions from interactions of
  the singlet and temporal sector are further suppressed, and omitted in this work. 
}
and only 
the ultrasoft fields remain.
These include fields driving
the transition (doublet and singlet) and spatial gauge fields. 

The dimensionally reduced high-temperature effective theory for the cxSM is parametrically of the same form as its 4d parent theory in eq.~\eqref{eq:lag-4d}.
We do not restate
the scalar Lagrangian for the doublet and singlet fields that defines the 3d EFT,
but merely label its parameters with the subscript 3.
Quartic couplings in the 3d EFT have dimension of mass ($T$), and
fields have dimension $T^{\frac{1}{2}}$.
We present an effective theory valid at next-to-leading order in dimensional reduction, 
namely
one-loop in couplings and fields and
two-loop for masses (except one-loop for parameter $b_{1,3}$).
This corresponds to the formal power counting
\begin{align}
\lambda_h^{ },
d_2^{ },
\delta_2^{ }
\sim g^2
\;,
\qquad
  \mu^2_h, b_2^{ }, b_1^{ }
\sim (g T)^2
\;,
\end{align}
that is accurate to $\mathcal{O}(g^4)$.
In the absence of cubic couplings,
the tadpole parameter ($a_{1,3}$) does not receive any loop corrections.
Accuracy higher than $\mathcal{O}(g^4)$, namely $\mathcal{O}(g^6)$, would require
additional higher dimensional operators~\cite{Kajantie:1995dw}.

The derivation of the dimensional reduction matching relations for generic models --
i.e. presenting 3d parameters as a function of temperature and parameters of parent theory --
has been demonstrated for example in~\cite{Kajantie:1995dw} in particular
for the SM and
for a real singlet in a recent tutorial~\cite{Schicho:2021gca}
(cf.\ also~\cite{Gould:2021dzl,Niemi:2021qvp}).
By extending this computation to the case of a complex singlet,
we derive the matching relations presented in appendix \ref{sec:DR}.
We have explicitly verified their gauge invariance:
matching relations at $\mathcal{O}(g^4)$ are independent of the gauge fixing choice of 
the parent 4d theory. 
We emphasise,
that in general dimensional reduction at NLO
(and at LO) is a gauge-invariant by construction;
see~\cite{%
  Jakovac:1994xg,Farakos:1994kx,Kajantie:1995dw,Croon:2020cgk,Schicho:2021gca,
  Hirvonen:2021zej}. 
For the past two decades,
the popularity of the 3d EFT approach for the EWPT phase transition thermodynamics has 
been superseded by that of lower order computations.
One presumable bottleneck denying more applications, is
the inability to perform dimensional reduction at NLO.
In this regard, we advocate upcoming software~\cite{Ekstedt:2022xxx} that
helps to avoid biting the bullet and
automates such computations for user-defined models. 

To capture non-perturbative IR effects related to the ultrasoft scale, lattice Monte Carlo simulations of
the 3d EFT~\cite{Farakos:1994kx,Kajantie:1995kf,Gould:2021dzl} are required.
For the cxSM, these simulations are out of scope of this work.
The remainder of the article,
concentrates on the perturbative computation of thermodynamics.
While inferior, and crucially lacking even
a qualitatively correct description for certain cases,%
\footnote{
  In crossover transitions all
  thermodynamic observables
  remain continuous, in particular derivatives of the order parameter.
  Such transitions are indiscernible in perturbation theory
  wherein a barrier always exists.
  This barrier -- albeit small -- then indicates a first-order transition. 
}
these perturbative studies are computationally much less expensive and
can be used as a first approximation and valuable guidance for future simulations. 
Nonetheless,
the 3d EFT mapping presented in appendix~\ref{sec:DR} is one of
the key ingredients for such simulations.
In addition, one needs lattice-continuum relations
(cf.~\cite{Laine:1995np,Laine:1997dy}),
simulation code including proper Monte Carlo update algorithms
(for a recent application, see~\cite{Gould:2021dzl} and references therein), and
a computation cluster.
 
%
\subsection{Thermal effective potential in perturbation theory}

The central quantity for equilibrium thermodynamics is
the free energy of the system, or the effective potential for
homogeneous field configurations.
We begin by recalling the perturbative expansion of this quantity, along
the lines of~\cite{Gould:2021oba} (also cf.~\cite{Farakos:1994kx}, and further e.g. \cite{Kajantie:2002wa,Kajantie:2003ax,Gynther:2005av,Gynther:2005dj,Laine:2015kra}
in the context of the QCD and electroweak pressure).

%
\subsubsection*{Perturbative expansion}

At zero temperature, the effective potential admits
a formal expansion in $g^2$, resulting in%
\footnote{
  For multiple couplings,
  one can organise perturbation theory by assigning each coupling a 
  formal power counting in $g$.
  In particular, we assume that the scalar quartic coupling scales as $g^2$.
}
\begin{align} 
\label{eq:VVacuumCouplingExpansion}
V^{T=0}_{\text{eff}} =
    \underbrace{A_2 g^2(\mu)}_{\text{tree-level}}
  + \underbrace{A_4[\ln \mu ] g^4}_{\text{1-loop}}
  + \underbrace{A_6 g^6}_{\text{2-loop}}
  + \ldots 
  \;.
\end{align}
We indicate the loop order at which each contribution arises.
In this case, the loop expansion aligns with the expansion in $g^2$.
Furthermore, we emphasised the {\em leading} dependence on
the renormalisation scale $\mu$ (in dimensional regularisation):
at tree-level an explicit dependence on this scale does not occur, but
couplings are implicit functions of $\mu$.
This scale dependence, or running, is governed by renormalisation group equations, or 
beta functions
$\mu \frac{{\rm d}}{{\rm d} \mu} g^2 \equiv \beta(g^2)$.
At one-loop order, these beta functions match
the coefficients of the logarithmic terms in
$A_4 \equiv A_4[\ln \mu]$
at one-loop level such that $\mu$-dependence cancels,
{\em viz.}
\begin{align}
  \LamD \frac{{\rm d}}{{\rm d}\LamD} V^{T=0}_{\rmii{eff}} \stackrel{\rmii{1-loop}}{=} \mathcal{O}(g^6)
\;.
\end{align}
This is the renormalisation group (RG) improvement at one-loop level.
It manifests when couplings are solved from their one-loop beta functions.
At higher loop levels, similar cancellations occur when
including higher-order beta functions and logarithmic terms. 
The key feature of the perturbative expansion at zero temperature is that
the full $\mathcal{O}(g^4)$ accuracy, with corresponding RG improvement,
can be achieved by a mere one-loop computation.
Furthermore, an error made by truncating at one-loop is parametrically of
$\mathcal{O}(g^6)$. 

The situation is more challenging in high-temperature perturbation theory.
Due to the enhancement of IR bosonic modes and subsequent resummations
in perturbation theory,
the formal expansion of the effective potential reads 
\begin{align} 
\label{eq:VThermalCouplingExpansion}
V^{\text{thermal}}_{\text{eff}} =
    \underbrace{a_2 g^2}_{\substack{\text{tree-level}\\\text{1-loop}}}
  + \underbrace{a_3 g^3}_{\text{1-loop}}
  + \underbrace{a_4 g^4}_{\substack{\text{1-loop}\\\text{2-loop}}}
  + \underbrace{a_5 g^5}_{\text{3-loop}}
  + \ldots
  \;.
\end{align}
The potential is
parametrically slower convergent than at zero temperature as
an expansion is in $g$ instead of $g^2$.
Hence,
to achieve the same accuracy as at zero-$T$ requires
the computation of higher loop levels.
In more detail
\begin{itemize}
\item[LO:]
  $g^2$-terms arise both at
    tree-level and from
    one-loop thermal corrections to masses.
\item[NLO:]
  $g^3$-terms arise solely at
    one-loop level in the 3d EFT for soft contributions.
\item[NNLO:]
  $g^4$-terms 
  comprise of:%
  \footnote{
    The zero-temperature Coleman-Weinberg potential at
    one-loop is included in (1) and (3). 
  }
\begin{itemize}
  \item[(1)]
    one-loop hard contributions in quartic terms,
  \item[(2)]
    one-loop field renormalisation contributions,
  \item[(3)]
    two-loop hard contributions to masses and
    one-loop mass corrections in the high temperature expansion,
  \item[(4)]
    two-loop soft contributions within the 3d EFT.
\end{itemize}
\item[$\text{N}^3$LO:]
  $g^5$-terms appear at three-loop level in 3d EFT~\cite{Rajantie:1996np}. 
\end{itemize}
The NLO one-loop term matches usual ``daisy resummation'' when only
the mass parameters are resummed by hard thermal corrections.
Practically,
in the 3d EFT approach also higher order resummations are
automatically included since also
couplings are resummed and
masses include two-loop corrections.
Truncated terms at $\mathcal{O}(g^6)$ include
two- and three-loop hard contributions,
contributions to higher dimensional operators in the 3d EFT, and
four-loop soft contributions within 3d EFT.%
\footnote{
  The four-loop soft contribution is already non-perturbative in
  non-Abelian gauge theories.
  Thus, an
  infinite number of 
  loops contribute~\cite{Linde:1980ts},
  rendering mere perturbation theory inconclusive at this order. 
}

Considering the above breakdown,
typical one-loop approximations of the thermal effective potential
are lacking in accuracy (cf.\ e.g.~\cite{%
  Delaunay:2007wb,Espinosa:2007qk,Profumo:2007wc,Noble:2007kk,Espinosa:2008kw,
  Espinosa:2011ax,Curtin:2014jma,Blinov:2015sna,Basler:2016obg,Basler:2017uxn,
  Chala:2018ari}).
They are only fully correct at $\mathcal{O}(g^3)$, and while they do include
a subset of
$g^4$-contributions related to one-loop potential (both zero-temperature and hard mode contributions), 
the computation is incomplete at that order.
It has an $\mathcal{O}(g^4)$ parametric error.
Such an inaccuracy is much worse than
the $\mathcal{O}(g^6)$ achieved at zero temperature at one-loop level. 
In particular, important logarithmic terms are missing at $\mathcal{O}(g^4)$
(such as those related to two-loop thermal mass) and
this causes a large residual RG scale dependence.
Using this feature one can probe
intrinsic, theoretical uncertainties in one-loop analyses of
the phase transition thermodynamics.
For some scenarios it was reported to be alarmingly large~\cite{%
  Kainulainen:2019kyp,Croon:2020cgk,Carena:2019une,Gould:2021oba}. 

A full $\mathcal{O}(g^4)$ accuracy is achievable
in the 3d approach with NLO dimensional reduction (two-loop for thermal masses) and 
within the 3d EFT perturbation theory at two-loop level;
cf.~\cite{%
  Farakos:1994kx,Kajantie:1995dw,Gorda:2018hvi,Kainulainen:2019kyp,Niemi:2018asa,
  Niemi:2020hto,Croon:2020cgk,Schicho:2021gca,Niemi:2021qvp,Gould:2021oba}.
In this case,
the error is of $\mathcal{O}(g^5)$ which is
still parametrically worse than one-loop level at zero temperature.
Only by including the three-loop effective potential within the 3d EFT,
one can reach 
$\mathcal{O}(g^6)$ accuracy --
the same as the one-loop accuracy at zero temperature.
For a real scalar field this has been computed in~\cite{Rajantie:1996np},
while for
the SM or its BSM extensions this computation remains outstanding.   
      
%
\subsubsection*{Radiatively generated barrier}

Before focusing on a concrete computation of the effective potential,
we comment on phase transitions with a radiatively generated barrier.
Since there a barrier is absent at
the tree-level potential (in 3d perturbation theory),
the barrier is provided by one-loop corrections.
Schematically, the one-loop potential for a generic 3d field $\Phi_3$ with
background field $\phi_3$ is of the form
\begin{align}
V^{\text{3d}}_{\text{eff}}(\phi_3) \simeq
    \frac{1}{2} \mu^2_3 \phi^2_3
  + \frac{1}{4} \lambda_3 \phi^4_3
  - \frac{1}{12\pi} \Big[
      \big( M^2(\phi_3) \big)^\frac{3}{2}
    + \big( m^2(\phi_3) \big)^\frac{3}{2}
  \Big]\;,
\end{align}
where the background field dependent mass eigenvalue
$m^2(\phi_3) \simeq \mu^2_3 + 3 \lambda_3^{ }\phi^2_3$
is related to the field $\Phi_3$ itself, and the mass eigenvalue $M^2$ is related to 
another, soft/heavy field with $M\sim g T$. 
The other field could be
a gauge field with
$M^2 \simeq (g_3 \phi_3)^2$ or
a second scalar with mass parameter
$\nu^2_3 \sim (g T)^2$ and
portal coupling
$a_{2,3} \sim g^2 T$ to the field $\Phi_3$, resulting in
$M^2 \simeq \nu^2_3 + \frac{1}{4} a_{2,3}^{ } \phi^2_3$.
For there to be a first-order phase transition
with different minima separated by a barrier at leading order, 
the cubic term has to be of same order as quadratic and quartic terms.
This can be achieved in the regime that admits a power counting
$\mu^2_3 \sim g^3 T^2$ and
$\lambda_3 \sim g^3 T$, resulting
in~\cite{Ekstedt:2020abj,Hirvonen:2021zej}
\begin{align}
V^{\text{3d,LO}}_{\text{eff}}(\phi_3) \simeq
    \frac{1}{2} \mu^2_3 \phi^2_3
    + \frac{1}{4} \lambda_3 \phi^4_3
    - \frac{1}{12\pi} \big( M^2(\phi_3) \big)^\frac{3}{2}
    \sim \mathcal{O}(g^3 T^3)
    \;.
\end{align}
Note,
that the term
$m^3 \sim g^{4.5} T^3$
is of higher order and does not appear in the leading-order potential. 
The formal perturbative expansion of
the effective potential for
a radiative barrier reads~\cite{Ekstedt:2022yyy,Gould:2022xxx} 
\begin{align}
\label{eq:radiative-barrier-expansion}
V^{\text{3d}}_{\rmii{eff}} 
\simeq
    \alpha_{3} g^{3}
  + \alpha_{4} g^{4}
  + \alpha_{4.5} g^{4.5}
  + \alpha_{5} g^{5}
  + \mathcal{O}(g^{5.5})
  \;.  
\end{align}
In this case,
convergence is even slower than in eq.~\eqref{eq:VThermalCouplingExpansion} and
the expansion is formally in $\sqrt{g}$ (instead of $g$).
In fact, there are two expansions:
one related to the heavy field,
where each higher contribution is suppressed by $g$ compared to previous order, and
one related to the light field
where each higher contribution is suppressed by $g^{1.5}$.
The latter expansion is responsible for non-integer powers in
eq.~\eqref{eq:radiative-barrier-expansion}.
Due to this structure of the expansion,
the $g^{3.5}$-term is absent. 
The NLO $g^4$-term arises from
two-loop contributions of heavy fields. 
The NNLO $g^{4.5}$-term arises at
one-loop order for this light field, as we have seen above.
The N$^3$LO $g^{5}$-terms are sourced from
three-loop diagrams of the heavy field. 

The work at hand performs computations in
the 3d EFT merely to one-loop order.%
\footnote{
  Crucially, dimensional reduction is still performed at two-loop level;
  see initial discussion of sec.~\ref{sec:setup}. 
} 
Consequently,
we can only access the leading-order potential for radiative transitions.

%
\subsubsection*{Computation of the effective potential}

Next, we compute the effective potential to one-loop order within
the 3d EFT~\cite{Farakos:1994kx}.
Doublet and singlet fields can be shifted by the real background fields
$v_3/\sqrt{2}$ and
$s_3/\sqrt{2}$,%
\footnote{
  For simplicity, we assumed that the imaginary component of the singlet
  does not admit a background field.
  The assumption can be relaxed for more general analyses.
}
respectively.
As a result the potential takes the form 
\begin{align}
\label{eq:veff-bg}
  V^{\rmii{eff}}_{\rmii{3d}}(v_3,s_3) =
  V^{\rmii{tree}}_{\rmii{3d}}(v_3,s_3)
  + \hbar V^{\rmii{1-loop}}_{\rmii{3d}}(v_3,s_3)
  + \mathcal{O}(\hbar^2)
  \;, 
\end{align}
which introduced $\hbar$ as a formal loop counting parameter within the 3d EFT.
Tree-level and one-loop contributions read
\begin{align}
V^{\rmii{tree}}_{\rmii{3d}}(v_3,s_3) &=
      \frac{1}{2} \mu^2_{h,3} v^2_3
    + \frac{1}{4} \lambda_{h,3}^{ } v^4_3
    \nn &
    + \sqrt{2}a_{1,3} s_3 
    + \frac{1}{2} \Bigl(\frac{b_{1,3}+b_{2,3}}{2}\Bigr)s^2_3
    \nn &
    + \frac{1}{16} d_{2,3}^{ } s^4_3
    + \frac{1}{8} \delta_{2,3}^{ } s^2_3 v^2_3
    \;, \\[2mm]
V^{\rmii{1-loop}}_{\rmii{3d}}(v_3,s_3) &=
     (d-1)\bigl(
      2 L_3(m_{\rmii{$W$},3})
    + L_3(m_{\rmii{$Z$},3})
    \bigr)
    \nn &
    + L_3(m_{\rmii{$A$},3})
    + L_3(m_{+,3})
    + L_3(m_{-,3})
    \nn &
    + 2\bigl( L_3(m_{2,+,3}) + L_3(m_{2,-,3}) \bigr)
    \nn &
    + L_3(m_{1,+,3})
    + L_3(m_{1,-,3})
    \;, 
\end{align}
where the one-loop master integral
(with \MSbar scheme dimensional regularisation) in
three dimensions has a simple result
\begin{align}
L_3(m) &\equiv
  \frac{1}{2} \Big( \frac{\Lamd^{2}e^\gamma}{4\pi} \Big)^\epsilon
  \int_{p} 
  \ln (p^2 + m^2) =
  - \frac{m^3}{12\pi}
  + \mathcal{O}(\epsilon)
  \;,
\end{align}
where
$\int_{p} = \int\frac{{\rm d}^{d}p}{(2\pi)^d}$
and 
the last equality holds for $d=3-2\epsilon$ dimensions.
Here
$\Lamd$ is the renormalisation scale of the 3d EFT and
$\gamma$ is the Euler-Mascheroni constant. 
The background field dependent mass eigenvalues
are functions of {\em 3d parameters} and
are collected in
appendix~\ref{sec:mass-eigenvalues}.
In the above one-loop expression,
we used general covariant, or Fermi, gauge with gauge parameters
$\xi_2$ for SU(2) and
$\xi_1$ for U(1) fields.
Gauge dependence appears only in
the Goldstone mass eigenvalues
$m_{1,3,\pm}$ in eq.~\eqref{eq:goldstone1} and
$m_{2,3,\pm}$ in eq.~\eqref{eq:goldstone2}.

The tree-level term captures the hard mode contributions at
$\mathcal{O}(g^2)$ and
$\mathcal{O}(g^4)$.
The one-loop term matches the conventional daisy resummed cubic terms
at $\mathcal{O}(g^3)$, while furthermore including
a subset of higher order resummations: 
all 3d 
parameters 
in $V^{\rmii{1-loop}}_{\rmii{3d}}$
are resummed at $\mathcal{O}(g^4)$, while
in typical LO daisy resummation only mass parameters are resummed,
at $\mathcal{O}(g^2)$. 
Importantly, in the 3d effective potential, via two-loop matching of
the mass parameters, we include the thermal masses and hence quadratic terms
at two-loop order.
In contrast, typical direct computations of the thermal effective potential only include these terms at one-loop.
Including two-loop thermal masses
improves 
the 3d EFT based approach by reducing renormalisation scale dependence,
as was pointed out in~\cite{Gould:2021oba} and
further demonstrated in sec.~\ref{sec:numerics}.   
For readers unfamiliar with the 3d EFT approach to resummation of
the thermal effective potential, see appendix A in~\cite{Gould:2021oba}
for a comparison to a typical thermal effective potential computed directly in
4d parent theories.  

Finally, we inspect 
the LO effective potential for 
a singlet field that undergoes the transition 
in the presence of parametrically heavier Higgs and gauge fields.
In practice,
this happens for the first step of a two-step phase transition.
By construction,
the heavy Higgs field does not acquire
a non-zero background field.
For simplicity, we assume here a $Z_2$-symmetric case,
such that the singlet tadpole does not enter.
The result reads
\begin{align}
\label{eq:veff-rad-LO}
V^{\rmii{eff,rb}}_{\text{3d}}(s_3) &= 
    \frac{1}{2} \Bigl(\frac{b_{1,3}+b_{2,3}}{2}\Bigr)s^2_3
  + \frac{1}{16} d_{2,3}^{ } s^4_3
  - 4\Big(\frac{(m^2_{\phi,3})^{\frac{3}{2}}}{12\pi}\Big)
  \;,
\end{align}
where
``rb'' stands for radiative barrier
and
$m^2_{\phi,3} \equiv \mu^2_{h,3} + \frac{1}{4} \delta_{2,3}^{ } s^2_3$
is the 4-degenerate doublet mass eigenvalue in
the presence of a non-zero singlet background field.
The last term is the one-loop contribution from the heavy Higgs,
that is formally $\mathcal{O}(g^3)$.
Note that gauge fields do not contribute at LO since
they do not couple to the singlet.
This result for the effective potential can be interpreted to correspond to
an alternative EFT where
the heavy Higgs field has been integrated out~\cite{Gould:2021ccf,Hirvonen:2021zej}.

%
\subsection{Thermodynamics}

The pressure encodes the information of equilibrium thermodynamics.
It is related to the effective potential as
$p =
- V^{\rmii{4d}}_{\rmii{eff}} =
- T V^{\rmii{3d}}_{\rmii{eff}}$ and
in particular the pressure differences between different phases that are described by the minima of the effective potential.  
At the critical temperature, the pressures of two phases are equal
$\Delta p = 0$ 
which translates to
the condition of degenerate minima in the effective potential 
$\Delta V_{\rmii{eff}}^{\rmii{3d}} = 0$. 
Here and below, we denote
$\Delta (\ldots) \equiv
(\ldots)_{\rmii{low}} -
(\ldots)_{\rmii{high}}$
for the difference between the low and high temperature phases. 
Hence, we do not consider those contributions to the pressure that are present when 
both background fields vanish~\cite{Gynther:2005av,Gynther:2005dj,Laine:2015kra},
since these are equal in the symmetric and broken phase. 
\begin{figure}[t]
 \centering
  \includegraphics[width=0.5\textwidth]{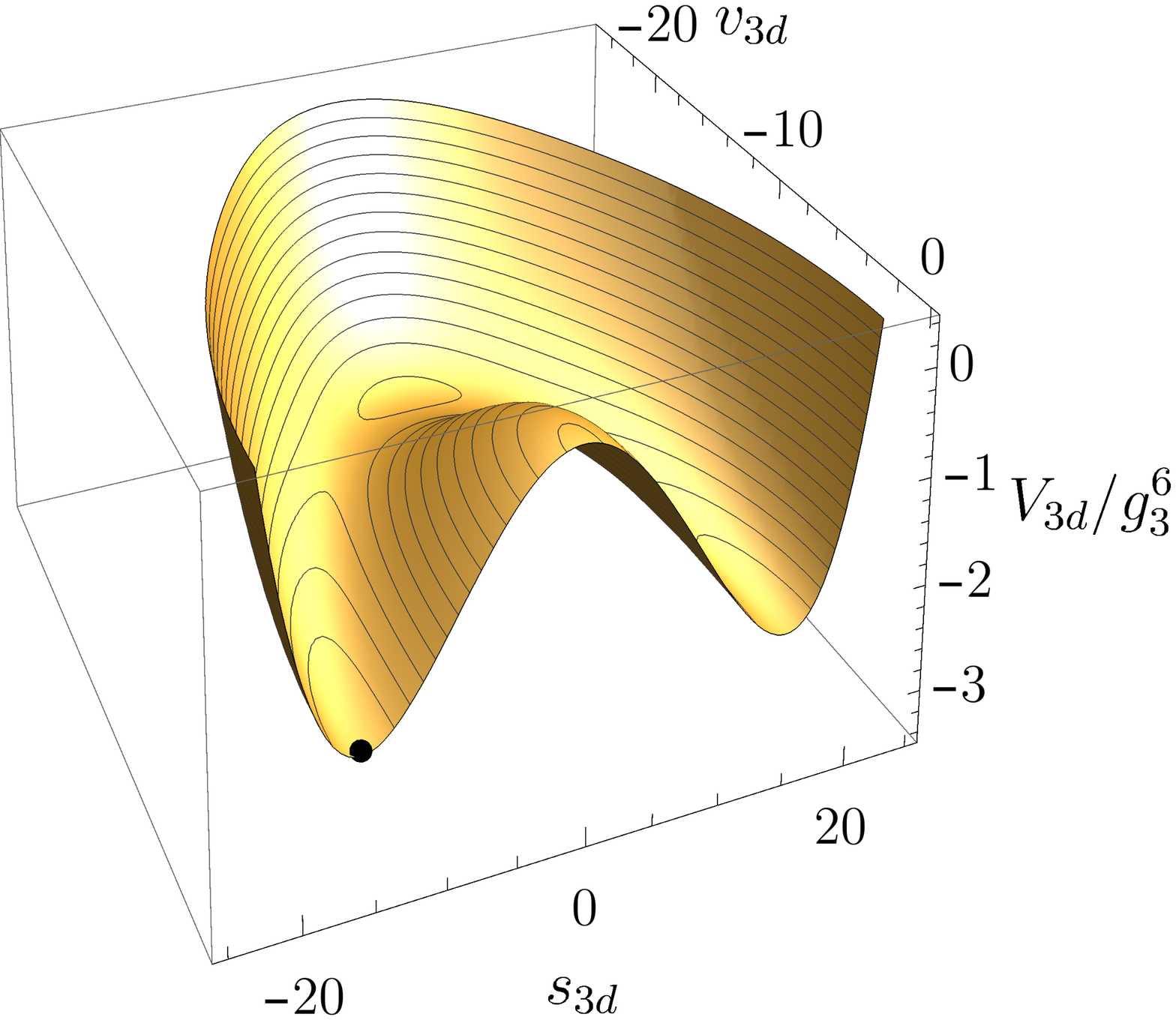}%
  \includegraphics[width=0.5\textwidth]{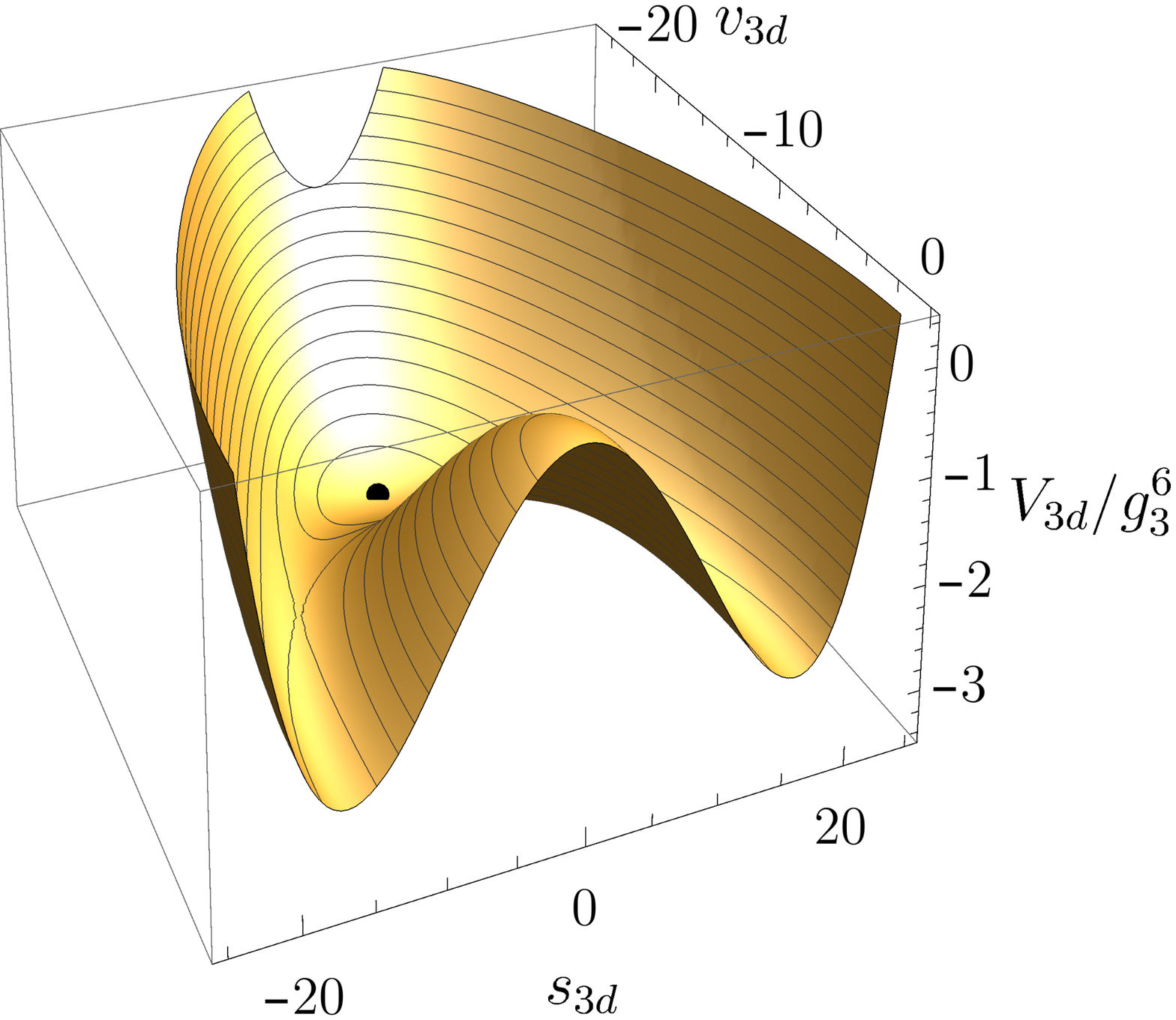} 
  \caption{%
    Schematic illustration for two-step phase transition, showcasing the effective potential near
    the critical temperature ($T_{\rmi{c},\phi}$) of
    the second phase transition from
    singlet to Higgs phase.
    At $T > T_{\rmi{c},\phi}$ (left),
    the minimum -- depicted with a black dot -- in singlet direction is global.
    At $T < T_{\rmi{c},\phi}$ (right),
    the Higgs phase becomes energetically favorable and
    the system undergoes a second phase transition.
    Since the singlet and Higgs minima are separated by a barrier,
    the transition is of first order.
    This barrier exists already at tree-level and therefore
    the second transition can be strong and relevant for
    gravitational wave production.
}
\label{fig:veff}
\end{figure}

A sketch of the effective potential for a two-step phase transition~\cite{Patel:2013zla,Inoue:2015pza,Niemi:2020hto,Bell:2020gug} is presented in
fig.~\ref{fig:veff}.
It focuses on the second transition from singlet to Higgs phase.
At higher temperatures,
the extremum at the origin becomes
the global minimum and
the system is in the symmetric phase.
In a first-order transition,
the scalar condensates defined as~\cite{Farakos:1994xh}
\begin{align}
\label{eq:condH}
\langle \phi^\dagger \phi \rangle &\equiv
  \frac{\partial V^{\rmii{eff}}_{\rmii{3d}}}{\partial \mu^2_{3}}
  \;, \\
\label{eq:condS}
\langle S^* S \rangle &\equiv
  2 \frac{\partial V^{\rmii{eff}}_{\rmii{3d}}}{\partial b_{2,3}}
  \;,
\end{align}
act in analogy to order parameters
since they 
can be discontinuous at the critical temperature.
The factor two in
eq.~\eqref{eq:condS} is a consequence of
the chosen normalisation in eq.~\eqref{eq:lag-4d} for the singlet mass term.
To measure the strength of the phase transition, we compute the latent heat
$L = T \Delta p'$.
Here, the prime denotes a temperature derivative.
In terms of the effective potential it reads 
\begin{align}
\label{eq:latent-heat}
L(T) &= T^2 \Delta \frac{{\rm d} V^{\rmii{eff}}_{\rmii{3d}}}{{\rm d}T}
\;.
\end{align}  
In a naive perturbative treatment, 
the effective potential is simply minimised at different temperatures to find
the phases, to determine
the critical temperature and strength of the transition, as described above.
However,
this treatment leads to a subtlety related to gauge invariance:
the value of the effective potential at its extrema, and therefore also at the minima, 
are gauge invariant.
However, the minima themselves, i.e.\ values of the background fields, are
gauge dependent.
Furthermore, outside the extrema,
the value of the effective potential is also gauge dependent and
a numerical minimisation to find the minima of the potential inherits
an artificial, residual gauge dependence of the gauge fixing parameters.
Therefore, care has to be invested into this issue, as done
in~\cite{Laine:1994zq,Patel:2011th,Ekstedt:2020abj}.

%
\subsubsection*{Gauge invariant computation}

The Nielsen identities~\cite{Nielsen:1975fs,Fukuda:1975di} guarantee that
the value of the effective potential in its minima are
gauge invariant order-by-order in perturbation theory.
In the $\hbar$-expansion~\cite{Laine:1994zq,Patel:2011th}
orders of perturbation theory are tracked down in powers of
a formal loop counting parameter.
Additionally,
the minima are expanded in $\hbar$ as:
$v^{\text{min}}_3 = v_{3,0} + \mathcal{O}(\hbar)$ and
$s^{\text{min}}_3 = s_{3,0} + \mathcal{O}(\hbar)$.
Here
$v_{3,0}$ and $s_{3,0}$ are 
the minima of the leading-order potential.
During a second step of a two step transition, these are simply minima of
the tree-level potential (cf.\ eqs.~\eqref{eq:min0}--\eqref{eq:minH}) and
the potential evaluated at the minima expands as 
\begin{align}
\label{eq:veff-hbar}
V^{\rmii{eff}}_{\rmii{3d}}(v^{\rmii{min}}_3,s^{\rmii{min}}_3) &= 
    V^{\rmii{tree}}_{\rmii{3d}}(v_{3,0},s_{3,0})
  + \hbar \; V^{\rmii{1-loop}}_{\rmii{3d}} (v_{3,0},s_{3,0})
  + \mathcal{O}(\hbar^2)
  \;.
\end{align}  
The generic form of the $\mathcal{O}(\hbar^2)$ correction can be found
in~\cite{Niemi:2020hto}, but we do not include it in our analysis as it requires
a two-loop computation of
$V^{\rmii{eff}}_{\rmii{3d}}$.
At $\mathcal{O}(\hbar^2)$ there would be additional contributions involving
derivatives of the tree-level and one-loop pieces with respect to
the background fields, as well as the two-loop potential itself. 
As we truncate our computation at $\mathcal{O}(\hbar)$,
the only difference to the effective potential in terms of generic background fields is 
that we evaluate both tree-level and one-loop parts at the tree-level minima.
In Fermi gauge and at one-loop level,
the gauge fixing parameters appear solely in Goldstone mass eigenvalues
$m_{1,\pm,3}$ and
$m_{2,\pm,3}$
(eqs.~\eqref{eq:goldstone1} and \eqref{eq:goldstone2}).
These vanish at the tree-level minima $v_{3,0},s_{3,0}$ which
in turn provides a gauge-invariant treatment at $\mathcal{O}(\hbar)$. 

The procedure described above is improved compared to 
the PRM-scheme proposed in~\cite{Patel:2011th}.
It consistently resums hard thermal loops (in 3d EFT parameters) to
next-to-leading order
(i.e.\ $\mathcal{O}(g^4)$ in a formal power counting in $g$) while
maintaining the gauge invariance.
As described earlier, this ensures partial RG improvement related to
the hard thermal scale with consistent resummation, and
reduces the intrinsic uncertainty of the computation~\cite{Croon:2020cgk,Gould:2021oba}.
In practice, we can find gauge-invariant critical temperatures in analogy
to~\cite{Patel:2011th}: 
by determining values of the effective potential of eq.~\eqref{eq:veff-hbar} in each
minimum as function of temperature, and
determining when the curves intersect 
(cf.\ fig.~\ref{fig:phases-condensates}~(left)).
On algorithm level~\cite{Patel:2011th},
this is more efficient than
the numerical minimisation of a complicated two- (or multi-) variate function. 
We also emphasise, that the condensates in
eqs.~\eqref{eq:condH}--\eqref{eq:condS} are gauge invariant
when evaluated using eq.~\eqref{eq:veff-hbar}.

While gauge invariance is manifest in this treatment~\cite{Laine:1994zq},
one subtlety remains:
radiatively generated or loop-induced transitions require additional care.
In fact, for an expansion in $\hbar$ to be meaningful,
the leading order minima
$v_{3,0}$ and/or
$s_{3,0}$ have to exist.
For the $Z_2$-symmetric case
(without mixing between scalar mass eigenstates)
tree-level broken minima exist only if the (3d EFT) mass parameters become negative.
In practice~\cite{Niemi:2020hto} this leads to a condition that
the broken minima are global minima immediately when these mass parameters change sign. Therefore in this approach, solutions for
the critical temperature are determined only from the condition that
the corresponding 3d EFT mass parameter vanishes.
This is not the physically correct picture.
Another reason to be cautious,
is that at two-loop order,
the effective potential, or its derivatives related to gauge-invariant condensates,
have a spurious IR divergence at a vanishing mass parameter
which renders the computation further non-predictive~\cite{Niemi:2020hto}.
Similarly, in the original ref.~\cite{Laine:1994zq}, an expression for
the critical temperature itself was
given in $\hbar$-expansion and
found to diverge at $\mathcal{O}(\hbar^2)$.
Here,
this problem 
haunts one-step transitions 
from the symmetric to the Higgs phase, as well as
the first step of a two-step scenario
from the symmetric to the singlet phase.
For the latter, we further demonstrate this issue below for
a concrete numerical example.

A cure of this technical problem is the resummation of
a subset of one-loop contributions to the leading-order potential.
Contributions from the heavy field give rise to a barrier at leading order
(cf.\ eq.~\eqref{eq:veff-rad-LO}).
Consequently, minima of the leading-order potential are away from
the point where the 3d EFT mass parameters vanish and
the problematic IR behaviour described above can be avoided.
However, as we only work with the leading-order potential for
a radiatively generated first step of the transition,
we are not able to directly demonstrate if higher order corrections are
indeed free from spurious IR divergences.
A related and detailed discussion on this topic can be found 
in~\cite{Gould:2022xxx}.

%
\section{Numerical analysis}
\label{sec:numerics}

To demonstrate numerically the setup for thermodynamics described in
the previous section,
we investigate a single parameter space point:
\begin{align}
\label{eq:input1}
\Bigl\{
  \frac{m_{\rmii{$H$}_2}}{\text{GeV}},
  \frac{\mA}{\text{GeV}},
  \delta_2, d_2
\Bigr\} = \{62.5, 62.5, 0.55, 0.5 \}
\;, 
\end{align}
with
$m_{\rmii{$H$}_1} = 125.1$~GeV identified with the observed Higgs boson mass.
The pole masses of the other scalar bosons,
$m_{\rmii{$H$}_2}$ and
$\mA$, are related to \MSbar parameters in appendix~\ref{sec:MSbars}, and the
singlet portal coupling and
self-interaction coupling are treated as input parameters.
This parameter space point matches the S2 scenario studied in~\cite{Chiang:2017nmu}, 
and admits a two-step phase transition scenario. 
In fact, this point belongs to
a subset of parameter space for which
the tree-level potential eq.~\eqref{eq:lag-4d} is
$Z_2$-symmetric $S\to -S$, {\em viz.}\
$a_1$ vanishes.
In the same parameter space point also
both the singlet VEV, $v_{S_0}$, and
the mass parameter $b_1$ vanish.
This simplifies the expressions in appendix~\ref{sec:MSbars}.
For the replicability of the analysis,
we explicitly write the \MSbar parameters at the initial scale $\MZ$: 
\begin{align}
\{\gY^2, \gs^2, g^2, {g'}^2 \} &= \{0.98, 1.48, 0.42, 0.12 \}
\;, \\
\Bigl\{
  \frac{\mu^2_h}{\text{GeV}^2},
  \frac{b_2}{\text{GeV}^2},
  \frac{b_1}{\text{GeV}^2},
  \frac{a_1}{\text{GeV}^3}
\Bigr\} &=
\{-7825, -8859, 0, 0 \}
  \;, \\[2mm]
\{\lambda_h, \delta_2, d_2 \} &= \{0.13, 0.55, 0.5\}
  \;,
\end{align}
which
are obtained by solving the \MSbar parameters as
a function of the input parameters of eq.~\eqref{eq:input1}
using the relations of appendix~\ref{sec:MSbars}.
Here,
we displayed rounded up numbers while our analysis uses higher decimal accuracy.

For comparison,
we describe below both the
gauge in- and dependent determinations of our numerical analysis.
It proceeds in the following steps:
\begin{itemize}
\item[1.]
  For a fixed input parameter space point,
  solving the \MSbar parameters by tree-level relations
  (cf.\ appendix~\ref{sec:MSbars}).
\item[2.]
  Solving
  RG running of the \MSbar parameters from
  the one-loop beta-functions (cf.\ appendix~\ref{sec:betas}).
\item[3.]
  For fixed temperature $T$,
  the parameters of the 3d EFT are obtained from
  the matching relations of appendix~\ref{sec:DR}
  in terms of the \MSbar parameters that are run to a chosen $T$-dependent scale.  
\item[4.]
  For fixed temperature $T$,
  the thermal effective potential within the 3d EFT is constructed as described 
  in the previous section either in
  the gauge-invariant $\hbar$-expansion at leading order minima, or in
  Fermi gauge as a function of generic background fields.   
\item[5.]
  Steps 3 and 4 are repeated for different values of $T$ and
  critical temperatures $\Tc$,
  condensates and latent heat are determined,
  as described in the previous section.
  We further demonstrate this in figures below in this section.
\end{itemize}
 
Partial RG improvement (related to the hard thermal scale) manifests through
steps 2 and 3, when the 4d RG scale cancels at $\mathcal{O}(g^4)$. 
Unlike in~\cite{Chiang:2017nmu}, where running is only implemented to
the tree-level part of the effective potential,
we use running couplings consistently everywhere.
This will induce contributions that are formally of higher order than
the accuracy of our computation.
In spirit of~\cite{
  Bodeker:1996pc,Laine:2017hdk,Croon:2020cgk,Gould:2021oba,
  Papaefstathiou:2020iag,Papaefstathiou:2021glr},
this indicates the intrinsic uncertainty of our analysis.
We emphasise that,
in a consistent cancellation of 4d RG scale at $\mathcal{O}(g^4)$,
two-loop pieces of 3d mass parameters --
or two-loop thermal masses --
are essential~\cite{Gould:2021oba}.%
\footnote{
  This feature is not only related to BSM physics but is present already in the SM
  as detailed in appendix~\ref{sec:light-higgs}.
}

This technical detail has been overlooked by all existing literature on
the cxSM thermodynamics.
Our computation is the first one to account and demonstrate
the importance of this partial RG improvement.
By working at $\mathcal{O}(\hbar)$ or one-loop within the 3d perturbation theory,
we do not have full, consistent leading RG improvement at $\mathcal{O}(g^4)$.
However, we can
estimate the magnitude of
the missing two-loop contributions by
including the 3d running of mass parameters and
varying the corresponding 3d RG scale.

The gauge invariance of our analysis is guaranteed by
the gauge-invariant matching relations of step 3 and
the $\hbar$-expansion in step 4.
For comparison, we also perform the gauge-dependent analysis in the Landau gauge
which has been one of the most common choices in the literature for
the EWPT in different BSM extensions. 

%
\subsubsection*{Critical temperature}

We first determine the different phases as a function of temperature
from the value of the 
effective potential at different local minima of
the leading-order potential.
For our benchmark point,
the formulae of the tree-level minima admit a simple closed form:
\begin{align}
\label{eq:min0}
\text{symmetric phase:}
  &\qquad
  \Bigl\{
    v_{3,0} = 0,
    s_{3,0} = 0
  \Bigr\}
\;, \\[2mm]
\label{eq:minS}
\text{singlet phase:}
  &\qquad
  \Bigl\{
    v_{3,0} = 0,
    s_{3,0} = -i \; \sqrt{\frac{2b_{2,3}}{d_{2,3}}}
  \Bigr\}
  \;, \\
\label{eq:minH}
\text{Higgs phase:}
  &\qquad
  \Bigl\{
    v_{3,0} = -i \; \sqrt{\frac{\mu^2_{h,3}}{\lambda_{h,3}}},
    s_{3,0} = 0
  \Bigr\}
  \;.
\end{align}
In other parameter space points, that are not considered here,
solutions for minima can be functionally more complicated.
In terms of the above expressions,
we denote
\begin{align}
V^{\text{sym}}_{\text{eff}} &\equiv V^{\text{3d}}_{\text{eff}}(0,0)
  \;, \\[2mm]
V^{S}_{\text{eff}} &\equiv
  V^{\text{3d}}_{\text{eff}}\biggl(0,-i\sqrt{\frac{2b_{2,3}}{d_{2,3}}}\biggr)
  \;, \\
V^{\phi}_{\text{eff}} &\equiv
  V^{\text{3d}}_{\text{eff}}\biggl(-i\sqrt{\frac{\mu^2_{h,3}}{\lambda_{h,3}}},0\biggr)
  \;.
\end{align}
\begin{figure}[t]
  \centering
  \includegraphics[width=0.5\textwidth]{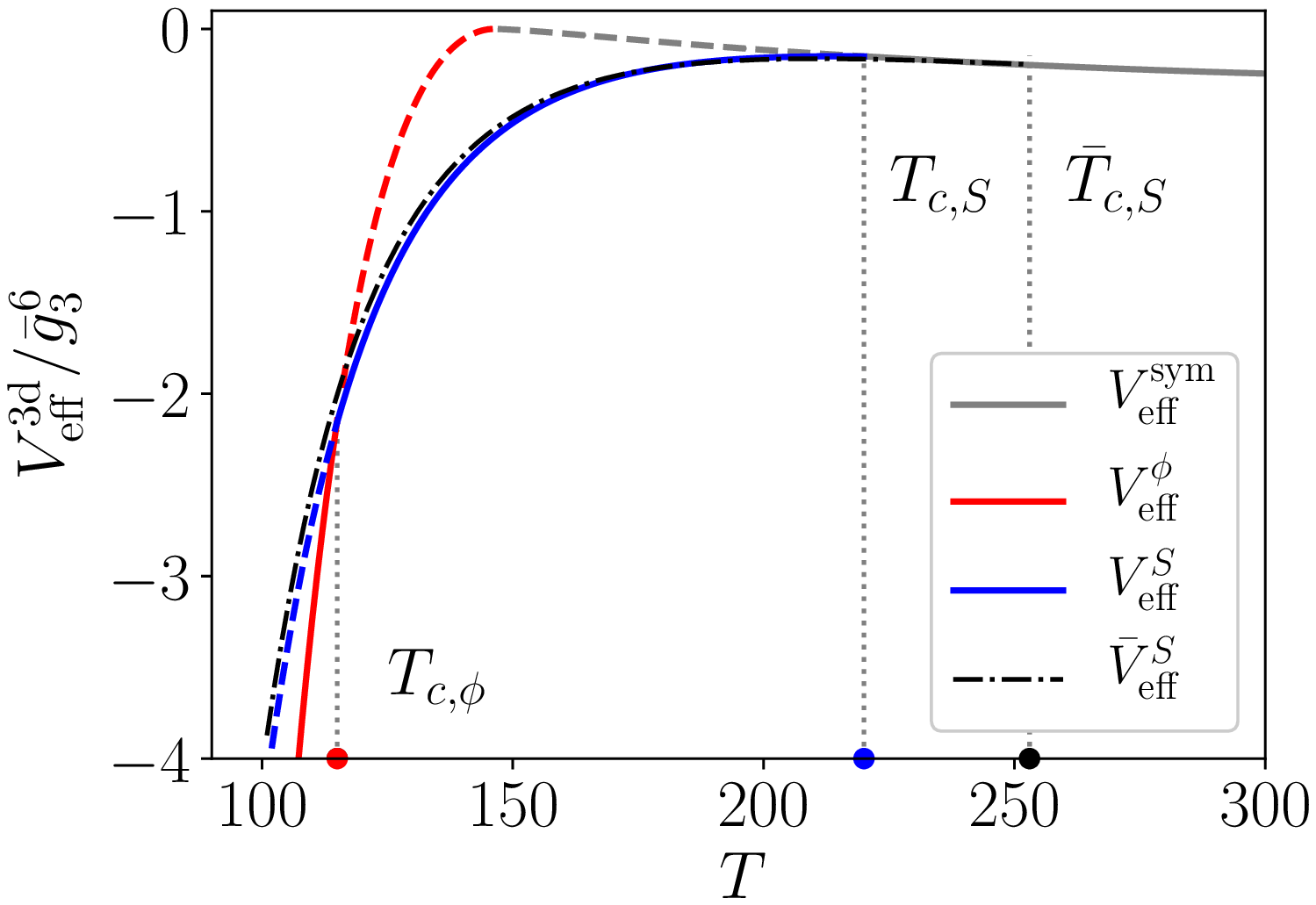}%
  \includegraphics[width=0.5\textwidth]{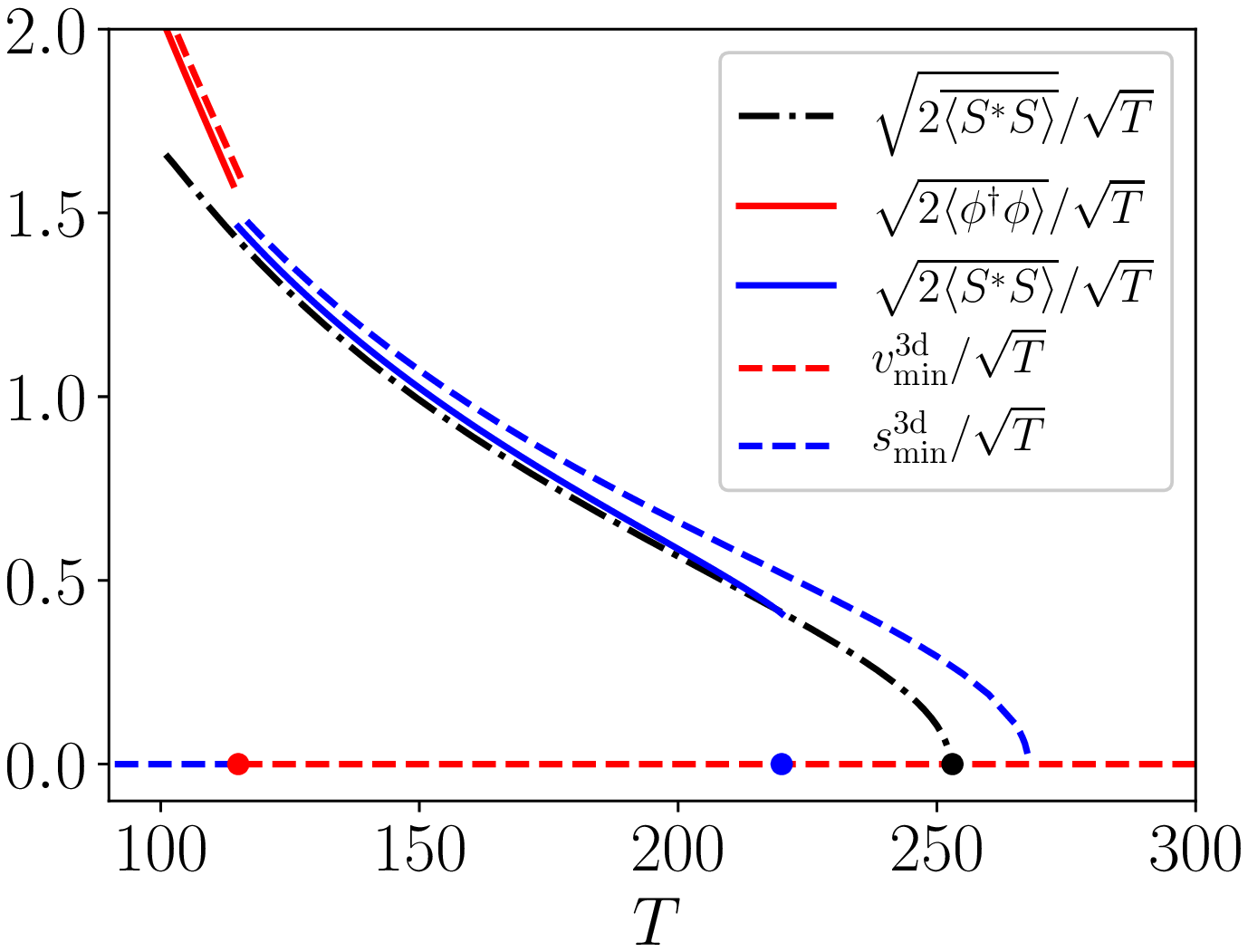} 
  \caption{%
    Left:
    Value of the effective potential as function of temperature $T$ in the
    symmetric (grey),
    Higgs (red) and
    singlet (blue) phase.
    Solid lines illustrate the global minimum.
    For dashed lines
    the potential also develops an imaginary part, signalling that
    the corresponding phase is not a stable minimum.
    Each phase is present only at limited $T$-intervals.
    Outside these ranges,
    the value of the effective potential is purely imaginary.
    Critical temperatures are determined from the intersection points,
    illustrated by
    dotted vertical lines and
    dots on the $x$-axis.
    The dotted-dashed black line describes
    the singlet phase in the EFT setup with a radiative barrier from the heavy Higgs.
    Right: 
    The solid blue and red lines show the square root of
    the gauge-invariant singlet and Higgs condensates
    in a temperature range for which they are real.
    The dotted-dashed black line presents
    the singlet condensate, in heavy Higgs EFT, in analogy to the left panel.
    For comparison,
    the dashed lines show the gauge-dependent values of the background fields
    in the global minima of the Landau gauge effective potential.
}
\label{fig:phases-condensates}
\end{figure}%
The left panel of fig.~\ref{fig:phases-condensates} plots these expressions as
a function of the temperature (cf.\ similar fig.~3 in~\cite{Patel:2011th}).
In this figure,
we present the value of the effective potential in
units of $g^6_3$;
the gauge coupling squared has dimension of mass in the 3d EFT.
Further, we used fixed RG scales
$\LamD = 1.25 \pi T$ and
$\Lamd = g^2_3$.
Later in this section,
we investigate how results change by varying these arbitrary RG scales.

The intersection points of
$V^{\text{sym}}_{\text{eff}}$,
$V^{S}_{\text{eff}}$, and
$V^{\phi}_{\text{eff}}$
determine critical temperatures.
On algorithmic level,
finding intersections of these curves is
significantly faster than minimising the (gauge-dependent) effective potential
which is a function of two variables.%
\footnote{
  We do not expand the critical temperature in $\hbar$ as in~\cite{Laine:1994zq}.
  Instead we numerically determine intersection points
  of the effective potential in different phases.
}
Since $V^{\text{sym}}_{\text{eff}}$ and
$\bar{V}^{\text{sym}}_{\text{eff}}$
are numerically indiscernible within the chosen plot ranges,
we do not visualise $\bar{V}^{\text{sym}}_{\text{eff}}$.

The global minimum is easily identified with the lowest value and
we denote it by
a solid line, while
dashed lines indicate that each local minimum is no longer the global one.
Each of the three minima are global in turn,
signalling a two-step phase transition. 
The intersection points of the global minima reveal the critical temperatures -- 
illustrated by vertical lines --
for both phase transitions. 
As already discussed above,
the determination of the critical temperature of the first transition, $T_{\rmi{c},S}$, is cumbersome as only after this point (i.e.\ at lower temperature)
the singlet minimum exists
(at higher temperature $V^{S}_{\text{eff}}$ has an imaginary value), and
$T_{\rmi{c},S}$ is therefore determined by the condition
$b_{2,3}(T_{\rmi{c},S}) = 0$.
This would lead to an IR divergence for the singlet condensate already at
two-loop level~\cite{Laine:1994zq,Niemi:2020hto} (also cf.~\cite{Kripfganz:1995jx}),
and further signals that
interpreting this point as the physical critical temperature is incorrect.
The physically correct picture is provided by solving the broken singlet minimum 
$\bar{s}_{3,0}$ from eq.~\eqref{eq:veff-rad-LO}, i.e.\
\begin{align}
  \frac{{\rm d}}{{\rm d} s_3} V^{\text{eff,rb}} (s_3) = 0
  \;,
\end{align}
at $s_3 = \bar{s}_{3,0}$.
This equation has four different solutions for
the broken phase extrema $\bar{s}_{3,0} \neq 0$ and in the case of interest,
the broken minimum is described by
\begin{align}
 \bar{s}_{3,0} &= \frac{1}{2\pi \sqrt{2}}
  \Bigl[ \frac{1}{d^2_{2,3}} \Bigl( \delta^3_{2,3} - (4\pi)^2 (b_{1,3}+b_{2,3}) d_{2,3}
  \nn &
  - \delta_{2,3} \sqrt{
      \delta^4_{2,3}
    + (4\pi)^2 \bigl(
        - 2(b_{1,3}+b_{2,3}) d_{2,3} \delta_{2,3}
        + 4 \mu^2_{S,3} d^2_{2,3}
      \bigr)}
    \Bigr)
    \Bigr]^{\frac{1}{2}}
  \;.
\end{align}
For a radiatively generated barrier,
the functional forms for the different extrema become seemingly more complicated.
However, the minimum at each temperature can still be identified by 
a derivative test 
of a single-variable function. 
We define the notation
$
\bar{V}^{\text{S}}_{\text{eff}} \equiv
V^{\text{eff,rb}} (\bar{s}_{3,0})$
and 
$
\bar{V}^{\text{sym}}_{\text{eff}} \equiv
V^{\text{eff,rb}} (0)$.
The critical temperature
$\bar{T}_{\rmi{c},S}$ is determined from
the temperature value when
$\bar{V}^{\text{S}}_{\text{eff}}$ becomes real and
$\bar{V}^{\text{S}}_{\text{eff}} < \bar{V}^{\text{sym}}_{\text{eff}}$.
Since
$\bar{T}_{\rmi{c},S}$ is different from
$T_{\rmi{c},S}$ in which $\mu^2_{S,3} = 0$,
the former solution for the critical temperature does not necessarily lead to
spurious IR divergences for condensates or latent heat
at higher loop levels.

%
\subsubsection*{Gauge invariant condensates}

The gauge-invariant condensates are analogous to
order parameters such that for
a first-order phase transition they are discontinuous at the critical temperature.
As an analogy to a typical
gauge-dependent analysis in terms of
gauge-dependent background fields,
we investigate the expressions~\cite{Laine:2017hdk,Niemi:2021qvp}
\begin{align}
\label{eq:vphys}
\frac{v_{\text{phys}}}{T} \equiv \frac{\sqrt{2 \langle \phi^\dagger \phi \rangle}}{\sqrt{T}}
  \;, \\
\label{eq:sphys}
\frac{s_{\text{phys}}}{T} \equiv \frac{\sqrt{2 \langle S^*S \rangle}}{\sqrt{T}}
  \;.
\end{align}
The condensates have discontinuities at the critical temperatures, and
the Higgs (singlet) condensate is negative outside of the Higgs (singlet) phase.
Therefore these expressions are imaginary outside of their respective phase. 
In fig.~\ref{fig:phases-condensates}~(right) we plot
the expressions of eqs.~\eqref{eq:vphys} and \eqref{eq:sphys} as
a function of temperature (solid lines).
For the singlet condensate,
we plot the result computed both from
$V^{\text{S}}_{\text{eff}}$ and 
$\bar{V}^{\text{S}}_{\text{eff}}$.
For the latter, we denote
$ \overline{\langle S^*S \rangle} \equiv 2 \frac{\partial \bar{V}^{\text{S}}_{\text{eff}}}{\partial b_{2,3}}$ and 
witness that sufficiently below
$\bar{T}_{\rmi{c},S}$ these results overlap until
a second phase transition.
We fixed RG scales at
$\LamD = 1.25 \pi T$ and
$\Lamd^{ } = g^2_3$.

The same plot presents the gauge-dependent background fields at the global minima of 
the effective potential (eq.~\eqref{eq:veff-bg}) in
Landau gauge $\xi_1 = \xi_2 = 0$ (dashed lines).
From this comparison, we observe 
that outside of the critical temperature of the first transition,
both approaches agree rather well.%
\footnote{
  The difference around the first transition critical temperature is unsurprising.
  In our gauge-independent computation,
  the singlet one-loop loop diagrams are not included as they are parametrically of 
  higher order and
  we have included merely the leading-order potential.
}
We interpret this as an echo of common lore in the literature
(cf.\ e.g.~\cite{Bodeker:1996pc}) that the Landau gauge results --
albeit inherently gauge-dependent, and therefore unphysical --
are numerically close to gauge-invariant results, 
perhaps encouraging their use as a practical guide.

On purely theoretical grounds,
gauge-dependent predictions for physical quantities are 
unsatisfactory and inherently incorrect.
Hence, it is invaluable to develop and use
a theoretically sound technique to analyse phase transition thermodynamics.
The Landau gauge results have been hoped to be useful in practice,
with the expectation that the error 
related to unphysical gauge dependence is smaller than other 
uncertainties of the problem
(cf.\ discussion on varying renormalisation scales below).
However,
the gauge fixing parameter is still an arbitrarily valued input.
In practice,
larger values could lead to larger numerical errors 
additional to the theoretical blemish.

%
\subsubsection*{Latent heat and renormalisation scale dependence}

\begin{figure}[t]
  \centering
  \includegraphics[width=0.5\textwidth]{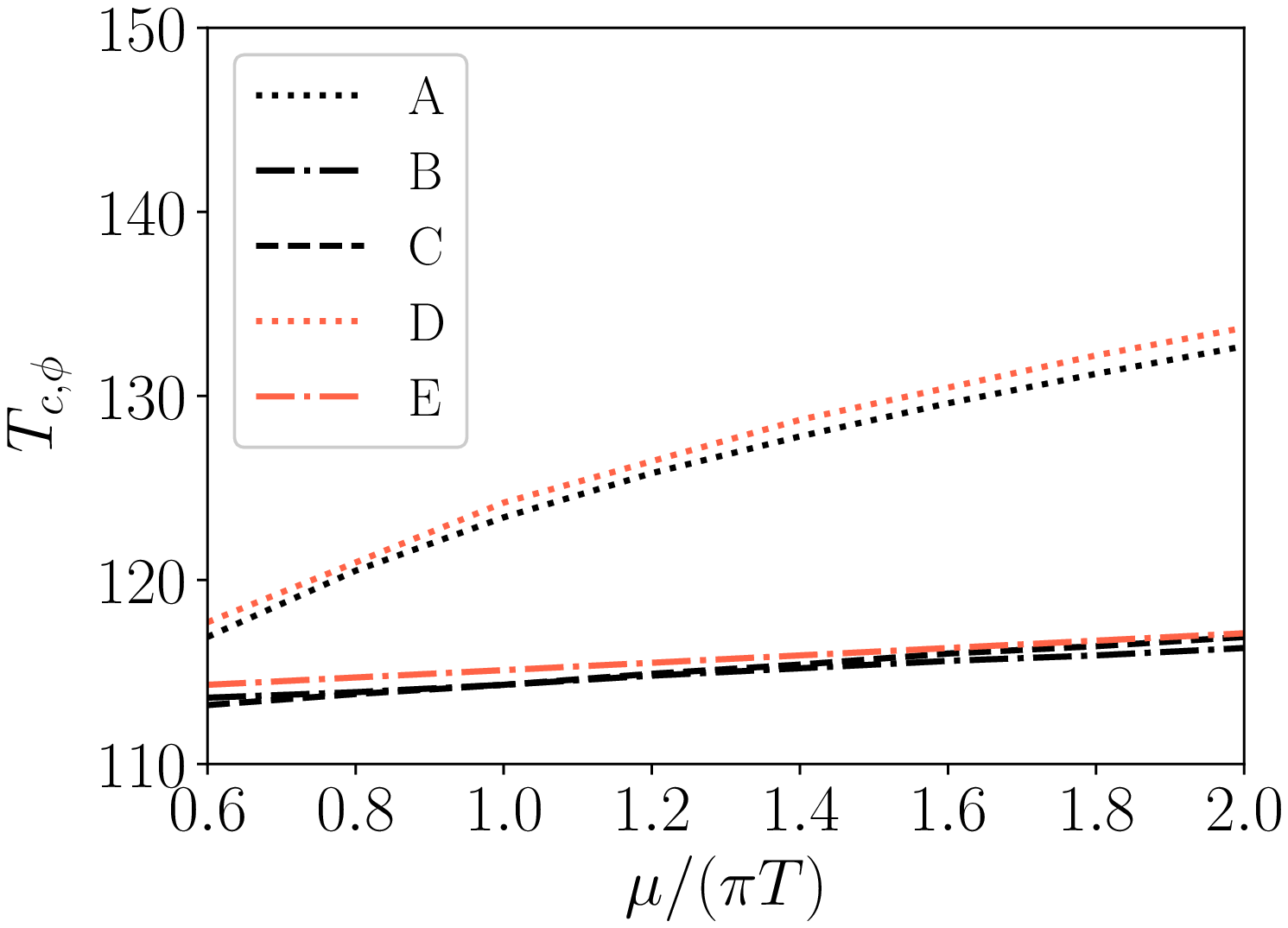}%
  \includegraphics[width=0.5\textwidth]{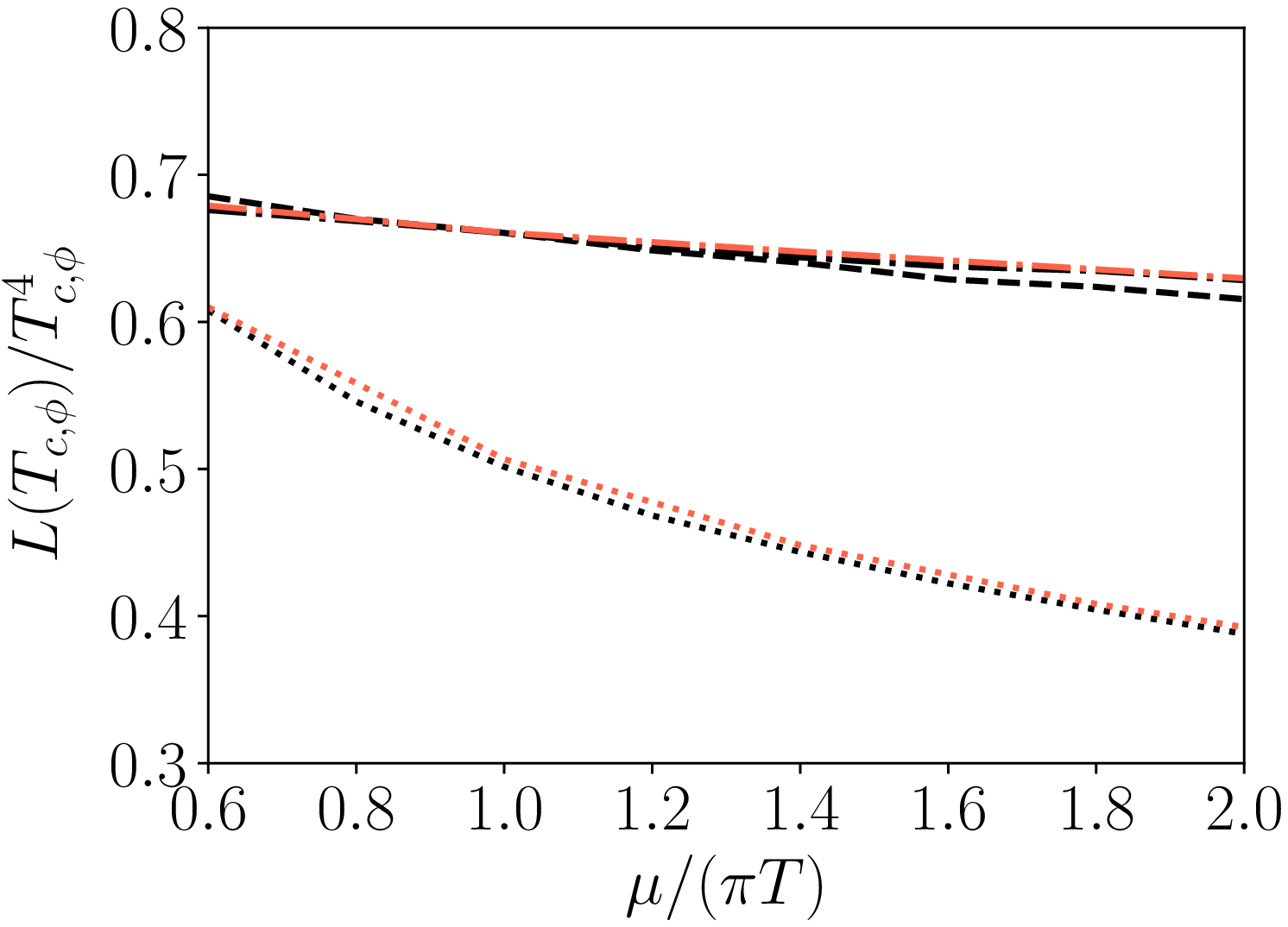} 
  \caption{%
    Left: $T_{\rmi{c},\phi}$ as function of
    the 4d renormalisation scale $\LamD$ in
    different approximations (A--E) that are detailed in the main body.
    The gauge-invariant $\hbar$-expansion (black lines A, B, C) and
    the analysis based on direct numerical minimisation of the effective potential in Landau gauge (grey lines D, E)
    show minor difference in the sensitivity to $\LamD$.  
    Right: As left panel, but for the latent heat evaluated at $T_{\rmi{c},\phi}$.
}
\label{fig:Tc-latent}
\end{figure}
The accuracy of our analysis is investigated
by varying the input renormalisation scales,
similar to recent studies~\cite{Croon:2020cgk,Gould:2021oba}.
The left panel of fig.~\ref{fig:Tc-latent} plots $T_{\rmi{c},\phi}$ --
the critical temperature of the second transition --
as a function of the 4d RG scale $\LamD$.
Similarly, fig.~\ref{fig:Tc-latent}~(right) evaluates 
the latent heat at $T_{\rmi{c},\phi}$.
We show five different lines:
lines A, B and C~(black) are based on the gauge-invariant $\hbar$-expansion while
lines D and E~(red) 
use a direct numerical minimisation of
the effective potential in Landau gauge.
Similar to fig.~\ref{fig:phases-condensates}~(right),
the two methods agree closely.
Concretely the lines contain:
\begin{itemize}
    \item [A,D:] one-loop level dimensional reduction,
    \item [B,E:] two-loop level dimensional reduction,
    \item [C:] as B, with varying $\Lamd = (\frac{\LamD}{\pi T}) g^2_3$.
\end{itemize}
The difference between the two 
levels of accuracy in dimensional reduction
arises from two-loop contributions to the thermal masses.
By comparison, we observe that 
without two-loop thermal masses, the results are strongly RG scale-dependent.
As reviewed in sec.~\ref{sec:setup},
the two-loop logarithmic terms compensate the leading running of
the one-loop thermal corrections to the mass parameters in
quadratic terms~\cite{Gould:2021oba}.
If these two-loop contributions are omitted,
the resulting uncertainty completely overshadows the ambiguity related to
the numerical difference of
gauge-invariant and
gauge-dependent (Landau gauge) approaches. 
Finally,
along the black dashed line C, we also varied the 3d RG scale $\Lamd$
(whereas for other lines it is fixed $\Lamd^{ }=g^2_3$).
This variation signals the importance of two-loop contributions at $\mathcal{O}(\hbar^2)$ within 3d perturbation theory that we omitted in
eq.~\eqref{eq:veff-hbar}. 
Note that mass parameters of the 3d EFT are running in terms of this 3d scale and 
logarithms of the 3d RG scale only appear at two-loop order.
While this effect is significantly smaller than two-loop contributions from
the hard thermal scale to the mass parameters,
it still causes a sizable uncertainty.
This motivates to increase the accuracy to
$\mathcal{O}(\hbar^2)$ in future computations.

%
\section{Discussion}
\label{sec:discussion}

The method presented in this article for phase transition thermodynamics avoids
the triune poison of
gauge dependence,
slow convergence of perturbation theory and
intractability.
Our investigation
follows~\cite{Laine:1994zq,Niemi:2020hto,Croon:2020cgk,Gould:2021oba} and
implements thermal resummations related to
the hard thermal scale of non-zero Matsubara modes using
the dimensionally reduced 3d EFT.
It further used
the gauge-invariant $\hbar$-expansion within 3d perturbation theory to 
compute the thermodynamic quantities pertinent to 
electroweak phase transitions.    
In analogy to previous studies~\cite{Croon:2020cgk,Gould:2021oba},
we find alarming sensitivity to the renormalisation scale at one-loop level and
we identified 
two-loop thermal masses to be the most crucial contributions to reduce this RG scale dependence.
Their importance was previously highlighted in~\cite{Gould:2021oba}.
We note that these findings are in contrast to~\cite{Cho:2021itv}.

Thus,
we suggest a minimal setup for perturbative accuracy, that
still eliminates most of the undesired scale dependence.
It combines
NLO dimensional reduction including two-loop thermal masses and 
a one-loop effective potential in the 3d EFT with a simple expression in terms of
the background field dependent mass eigenvalues.
The difference to~\cite{Niemi:2020hto,Croon:2019iuh,Niemi:2021qvp,Gould:2021oba} is 
that the two-loop effective potential in 3d EFT is not included;
the RG improvement related to hard thermal scale can still be acquired.%
\footnote{
  This omission is deliberate in favor of technical simplicity.
  Computing the effective potential in the broken phase 
  in terms of the background fields is
  significantly more complicated
  at two- than
  at one-loop level.   
  Instead, thermal masses for scalars can be computed in
  the unbroken phase from hard mode contributions to
  two-point correlation functions,
  and such a two-loop computation is standard in dimensional reduction literature.
  Nonetheless,
  the two-loop effective potential 
  would allow further RG improvement related to 3d EFT renormalisation scale.
}
The described setup relies on the ability to construct
the 3d EFT matching relations by dimensional reduction.
Such a setup is, unfortunately, still rare in current BSM EWPT literature.
To this end, we expect upcoming automated software~\cite{Ekstedt:2022xxx} --
designated for this problem --
to increase the applicability of improved studies based on the 3d EFT.
We further comment that ref.~\cite{Curtin:2016urg} has attempted
to improve resummation of hard thermal loops and in particular
to compute thermal masses beyond leading order based on 
``partial dressing''~\cite{Boyd:1993tz} but 
without using high-temperature expansion nor dimensional reduction to 3d EFT.
Another computation of the two-loop thermal effective potential
without high-temperature expansion appears
in~\cite{Laine:2017hdk} (cf.\ also~\cite{Laine:2000kv}).

The results of our gauge-invariant computation do not differ
drastically numerically from 
conventional Landau gauge analyses.%
\footnote{
  Our analysis is performed in one single benchmark point and
  the comparison concerns only Landau gauge.
  Therefore, we refrain from generic statements and
  acknowledge the theoretical importance of a gauge-invariant computation.
  In practice, most studies still ignore the issue of gauge dependence. 
}
Such a difference is completely overshadowed in a mere one-loop analysis
by the uncertainty from the renormalisation scale.
This, however, supports the common wisdom that while
a gauge-invariant computation is theoretically important, 
it should not come at the expense of
resummation and including relevant terms
in the coupling expansion.
In part, a similar conclusion was reached in~\cite{Ekstedt:2020abj}.
In this regard, our computation significantly improves
the previously suggested PRM scheme~\cite{Patel:2011th} by
incorporating the required resummations consistently,
while maintaining gauge invariance.

Finally, for the cxSM, 
the questions studied in~\cite{Chiang:2017nmu,Cho:2021itv}
could benefit from the tools presented in this article.
In particular, by focusing on what can be concluded about
the possible thermal history of EWSB in this scenario,
when both
gauge invariance and
RG improvement are installed.
The presented tools can be used for
wide scans of the model parameter space and 
to analyse physical implications. 
The latter can shed light on
which regions of parameter space admit both
a strong first-order phase transitions and dark matter candidates and
what are the collider phenomenology signatures of these regions. 
In addition to the improved equilibrium thermodynamics computation of this article,
future scans would benefit
from a two-loop effective potential in 3d perturbation theory, 
as well as
from the one-loop improved (zero-temperature) relations between
\MSbar parameters and physical input parameters.
Both of these improvements are available in
the real singlet-extended Standard Model (xSM)
collected in appendices A and B of~\cite{Niemi:2021qvp} and
could be generalised to the cxSM. 

A further computation of the bubble nucleation rate is required
for studying the gravitational wave production in the cxSM.
Therefore, one can combine
the 3d EFT presented in this article
with recent technology~\cite{Gould:2021ccf}.
In general, the same approach applies to other BSM theories and
in particular those with extended scalar sector.

%
\section*{Acknowledgments}
We thank 
Andreas Ekstedt,
Oliver Gould,
Joonas Hirvonen,
Thomas Konstandin,
Johan L{\"o}fgren, 
Lauri Niemi
and
Jorinde van de Vis 
for discussions.
PS has been supported
by the European Research Council, grant no.~725369, and
by the Academy of Finland, grant no.~1322507.
TT has been supported in part under
National Science Foundation of China grant no.~19Z103010239.
The work of GW is supported by World Premier International Research Center Initiative (WPI), MEXT, Japan.
%
\appendix
\renewcommand{\thesection}{\Alph{section}}
\renewcommand{\thesubsection}{\Alph{section}.\arabic{subsection}}
\renewcommand{\theequation}{\Alph{section}.\arabic{equation}}

%
\section{Collected formulae}
\label{sec:details}

This appendix collects several formulae that connect
the input parameters of
eq.~\eqref{eq:input1} to
the effective potentials in
eqs.~\eqref{eq:veff-bg} and \eqref{eq:veff-rad-LO}.
We also collect
numerical results for
the thermodynamics presented by the figures in sec.~\ref{sec:numerics}. 

%
\subsection{Relations between \MSbar parameters and input parameters}
\label{sec:MSbars}

In the gauge and Yukawa sector,
we use input values~\cite{ParticleDataGroup:2018ovx}
\begin{align}
\{ \MW,\MZ,M_t \} =
\{
  80.379~{\rm GeV},
  91.1876~{\rm GeV},
  172.76~{\rm GeV}
\}
  \;, 
\end{align}
together with
the strong coupling
$\gs = 1.48409$ 
and
the reduced Fermi constant
$\GF = 1.1663787 \times 10^{-5}~{\rm GeV}^{-2}$.
The strong coupling enters the two-loop thermal mass for Higgs doublet.
Using the shorthand notation
$g^2_0 \equiv 4\sqrt{2}\GF^{ }\MW^2$,
we have
\begin{align}
g^2 &= g^2_0
\;, \quad \quad
{g'}^2 = g^2_0\Big( \frac{\MZ^2}{\MW^2} - 1 \Big)
\;, \quad \quad
\gY^2 = \frac{1}{2} g^2_0 \frac{M^2_t}{\MW^2}
\;, \quad \quad
v_0 = \sqrt{\frac{4\MW^2}{g^2_0}}
\;.
\end{align}
To relate the \MSbar parameters to the above input parameters,
we use analytic, tree-level relations
\begin{align}
b_1 &= m^2_{\rmii{$H$}_2} - \mA^2
\;, & 
b_2 &= m^2_{\rmii{$H$}_2} + \mA^2 - \frac{1}{2} v^2_0 \delta_2
\;,\nn 
\mu^2_h &= - \frac{1}{2} m^2_{\rmii{$H$}_1}
\;, & 
  \lambda_h &= \frac{1}{2} \frac{m^2_{\rmii{$H$}_1}}{v^2_0}
\;. 
\end{align}
In many other parameter space points,
the relations between
\MSbar parameters and
input parameters do not have such simple analytic relations
for non-vanishing $a_1$, $v_{S_0}$ and
the mixing angle $\alpha$.
In these cases, one can solve the \MSbar parameters numerically by inverting
the mass eigenvalues
($m_+$ is identified with $m_{\rmii{$H$}_2}$ and
$m_-$ with $m_{\rmii{$H$}_1}$), 
together with the tadpole conditions
\begin{align}
v_0 \Big(
    \mu_h^2
  + \lambda_{h}^{ } v^2_0
  + \frac{1}{4} \delta_{2}^{ } v^2_{S_0}
  \Big) &= 0
\;, \\ 
  \sqrt{2} a_{1}^{ }
+ \frac{1}{2}(b_1 + b_2) v_{S_0}
+ \frac{1}{4} d_{2}^{ } v_0^3
+ \frac{1}{4} \delta_{2}^{ } v_0^2 v_{S_0}
  &= 0
\;,
\end{align}
and
the equation for the mixing angle
\begin{align}
\cot(2 \alpha) &= - \bigg(
  \frac{
    -4 \mu^2_h + 2(b_1 + b_2)
    + (3 d_2 - \delta_2) v^2_{S_0}
    + (\delta_2 - 12 \lambda_h) v^2_0
  }{4 \delta_2 v_0 v_{S_0}}
  \bigg)
  \;. 
\end{align}
All above relations hold at tree-level and receive quantum corrections in
zero-temperature perturbation theory.
In a consistent power counting up to $\mathcal{O}(g^4)$,
one would need to include
one-loop corrections to relations of \MSbar and physical parameters
(cf.~\cite{%
    Kajantie:1995dw,Laine:2017hdk,Niemi:2018asa,Kainulainen:2019kyp,Niemi:2021qvp}) --
we do not include them here.
Due to this omission 
we do not have the proper initial conditions
in our partial RG-improvement at $\mathcal{O}(g^4)$.
However, our discussion remains qualitatively intact as
these initial conditions merely shift
the corresponding benchmark point in the \MSbar parameter space.
Once this point is carefully related to particle collider phenomenology constraints, 
the improved initial conditions become quantitatively relevant to be accounted for.     

%
\subsection{Renormalisation group equations}
\label{sec:betas}

By parametrising the \MSbar renormalization scale through 
$t=\ln\LamD^2$
the one-loop $\beta$-functions, or renormalisation group equations,
for the \MSbar parameters read 
\begin{align}
\partial_{t}\lambda_{h} &=
    \beta_{\rmii{SM}}(\lambda_{h})
  + \frac{1}{(4\pi)^2} \Big( \frac{1}{8} \delta^2_2 \Big)
\;, \\
\partial_{t} d_{2} &=
  \frac{1}{(4\pi)^2} \Big( \frac{5}{2} d^2_2 + \delta^2_2 \Big)
\;, \\
\partial_{t} \delta_{2} &=
  \frac{1}{(4\pi)^2} \delta_2 \Big( d_2^{ }
  + \delta_2^{ }
  + 6 \lambda_h^{ } 
  - \frac{3}{4}  (3g^2 + {g'}^2)
  + 3\gY^2
  \Big)
\;, \\
\partial_{t} \mu^2_h &=
    \beta_{\rmii{SM}}(\mu^2_h)
  + \frac{1}{(4\pi)^2} \Big( \frac{1}{4} b_2 \delta_2 \Big)
\;, \\
\partial_{t} b_{2} &=
  \frac{1}{(4\pi)^2} \Big( b_2^{ } d_2^{ } + 2 \delta_{2}^{ } \mu_{h}^2 \Big)
\;, \\
\partial_{t} b_{1} &=
  \frac{1}{(4\pi)^2} \Big( \frac{1}{2}b_1 d_2 \Big)
  \;.
\end{align}
In the beta-functions for
$\mu^2_h$ and $\lambda_h^{ }$, 
we depict explicitly
only the new complex singlet contributions;
pure SM contributions are collected in e.g.~\cite{Brauner:2016fla}.  

%
\subsection{Mass eigenvalues for background field method}
\label{sec:mass-eigenvalues}

Parametrising doublet and singlet fields in analogy to eq.~\eqref{eq:fields}, but replacing zero-temperature VEVs,
$v_0$ and $v_{S_0}$ by generic, real background fields
$v$ and $s$ results in mass eigenvalues for the background field method
\begin{align}
m^2_\chi &= \mu^2_h + \lambda_h v^2 + \frac{1}{4} \delta_2 s^2, \\  
\mA^2 &= 
    \frac{1}{2} (-b_1 + b_2)
  + \frac{1}{4} d_2 s^2
  + \frac{1}{4} \delta_2 v^2
  \;, \\  
m^2_{\pm} &= \frac{1}{4} \bigg\{
    2 \mu^2_
  + b_1 + b_2
  + (6\lambda_h + \frac{1}{2}\delta_2 )v^2
  + \frac{1}{2}(\delta_2 + 3 d_2) s^2
  \nn &
  \pm \frac{1}{2} \sqrt{
    c_1
  + c_2 v^4
  + c_3 s^4
  + c_4 v^2
  + c_5 s^2
  + c_6 v^2 s^2}
\bigg\}
\;,
\end{align}
where we used the shorthand notation
\begin{align}
  c_1 &= 4 (b_1^{ } + b_2^{ } - 2 \mu^2_h)^2
  \;, \\
c_2 &= (\delta_2 - 12 \lambda_h)^2
  \;, \\
c_3 &= (\delta_2 - 3 d_2)^2
  \;, \\
c_4 &= 4 (\delta_2^{ } - 12 \lambda_h^{ })(b_1^{ } + b_2^{ } - 2 \mu^2_h) 
  \;, \\
c_5 &= -4(\delta_2^{ } - 3 d_2^{ })(b_1^{ } + b_2^{ } - 2 \mu^2_h)
  \;, \\
c_6 &= 6 d_2 (\delta_2 - 12 \lambda_h) + 2 \delta_2 (7 \delta_2 + 12 \lambda_h)
  \;.
\end{align}
These mass eigenvalues, in terms of parameters of
the 4d parent theory~\eqref{eq:lag-4d}, can be inverted to obtain
the tree-level relations between \MSbar parameters and physical masses in appendix~\ref{sec:MSbars}. 

When computing the effective potentials in the 3d EFT,
eqs.~\eqref{eq:veff-bg} and \eqref{eq:veff-rad-LO},
we denote background field dependent mass eigenvalues by
an additional subscript 3.
Then all parameters and background fields therein are those of the 3d EFT.
In addition,
when computing the effective potential in Fermi gauge, Goldstone mass eigenvalues $m^2_\chi$ are replaced by~\cite{Andreassen:2013hpa}
\begin{align}
\label{eq:goldstone1}
  m^2_{1,\pm} &= \frac{1}{2} \Big(
  m^2_\chi \pm
  \sqrt{m^2_\chi \big( m^2_\chi - v^2 (g^2 \xi_2+{g'}^2 \xi_1) \big)}
  \Big)
  \;, \\
\label{eq:goldstone2}
  m^2_{2,\pm} &= \frac{1}{2} \Big(
  m^2_\chi \pm
  \sqrt{m^2_\chi \big( m^2_\chi - v^2 g^2 \xi_2 \big)}
  \Big)
  \;,
\end{align}
where $m^2_{2,\pm}$ are double-degenerate.
The singlet does not contribute to
mass eigenvalues for weak bosons, {\em viz.}
\begin{align}
\mW^2 &= \frac{1}{4} g^2 v^2
  \;,\quad 
\mZ^2 = \frac{1}{4} (g^2 + {g'}^2) v^2
  \;,
\end{align}
and
the mass eigenvalue for the photon vanishes.

%
\subsection{Matching relations for dimensional reduction}
\label{sec:DR}

We define a shorthand notation
\begin{align}
L_{b} &\equiv
    2 \ln\Big( \frac{\mu}{T} \Big)
  - 2 \Big( \ln(4\pi) - \gamma \Big)
    \;, \\
c &=
  \frac{1}{2} \Bigl(
    \ln \Big( \frac{8\pi}{9} \Big)
    + \frac{\zeta_{2}'}{\zeta_{2}}
  - 2 \gamma \Bigr)
  \;,
\end{align}
where 
$\zeta_{s}$ for $\Re\,(s) > 1$ is the Riemann Zeta function.
Matching relations between parent 4d theory and 3d EFT read
\begin{align}
\label{eq:match:1}
\lambda_{h,3} &=
    \lambda_{h,3,\rmii{SM}}
  + T \frac{L_b}{(4\pi)^2} \Big( -\frac{1}{8} \delta^2_2 \Big)
  \;, \\
d_{2,3} &= T \Big(
      d_2(\mu)
    - \frac{L_b}{(4\pi)^2} \Big( \frac{5}{2} d^2_2 + \delta^2_2 \Big)
    \Big)
  \;, \\
\delta_{2,3} &= T \Big(
    \delta_2(\mu)
  - \frac{L_b}{(4\pi)^2}\delta_2^{ } \Big(
      d_2^{ }
    + \delta_2^{ }
    + 6 \lambda_h^{ }
    - \frac{3}{4}(3g^2 + {g'}^2)
    \Big)
    \nn &
    \hphantom{{}=T \Big(\delta_2(\mu)}
    - \frac{L_f}{(4\pi)^2} \delta_2^{ } \Big( 3 \gY^2 \Big)
    \Big)
  \;, \\
\mu^2_{h,3}(\Lamd) &= \mu^2_{h,3,\rmii{SM}}
    + \frac{1}{24}T^2 \delta_2(\LamD)
    - \frac{L_b}{(4\pi)^2} \Big( \frac{1}{4} b_2 \delta_2  \Big) 
    \nn &
    + \frac{1}{(4\pi)^2} T^2 \frac{\delta_2}{24} \Big(
        \frac{3}{4}(3g^2 + {g'}^2) L_b
        - 3\gY^2 L_f  \Big)
    \nn &
    - \frac{T^2 L_b}{(4\pi)^2} \delta_{2}\Big(
        \frac{1}{24} d_2
      + \frac{5}{48} \delta_2 
      + \frac{1}{4} \lambda_h  \Big)
    \nn &
    + \frac{1}{(4\pi)^2} T^2 \Big(
      - \frac{1}{4} \delta^2_2 \Big)
    \Big( c + \ln\Big( \frac{3T}{\Lamd} \Big) \Big)
    \;, \\
b_{2,3}(\Lamd) &= b_2(\LamD)
  + \frac{1}{6} T^2 \Big(d_2(\LamD) + \delta_2(\LamD) \Big)
  - \frac{L_b}{(4\pi)^2} \Big(
        b_{2}^{ }d_{2}^{ }
      + 2\delta_{2}^{ }\mu_{h}^{2} \Big)
  \nn &
  + \frac{T^2}{(4\pi)^2} \frac{\delta_2}{12} \Big( 3g^2 + {g'}^2 + 3 \gY^2 L_f \Big) 
  \nn &
  + \frac{T^2 L_b}{(4\pi)^2} \Big( 
    - \frac{5}{12} d^2_2
    - \frac{1}{6} d_2^{ } \delta_2^{ }
    + \frac{3}{8} g^{2}\delta_{2}^{ }
    + \frac{1}{8} {g'}^2 \delta_{2}^{ }
    - \frac{3}{4} \gY^{2}\delta_{2}^{ }
    - \frac{1}{3} \delta_{2}^{2}
    - \delta_{2}^{ } \lambda_h^{ }
  \Big)
  \nn &
  + \frac{1}{(4\pi)^2} T^2 \Big(
    - d^2_2
    - \delta^2_2
    + (3g^2 + {g'}^2 )\delta_2 \Big)
    \Big( c + \ln\Big( \frac{3T}{\Lamd} \Big) \Big)
    \;, \\
b_{1,3} &= b_1(\LamD)
  + \frac{L_b}{(4\pi)^2} \Big( -\frac{1}{2}b_1 d_2 \Big)
  \;, \\
\label{eq:match:7}
a_{1,3} &= \frac{1}{\sqrt{T}} a_1
  \;. 
\end{align}
Here,
we explicitly suppress
pure SM contributions since they can be found e.g.\
in~\cite{Schicho:2021gca} together with
the matching relations for the 3d gauge couplings and parameters of
the temporal sector.
The tutorial of~\cite{Schicho:2021gca} also
computes similar matching relations for a real singlet field.
The tadpole coupling $a_{1,3}$ does not receive loop corrections
since cubic couplings are absent.

All matching relations eqs.~\eqref{eq:match:1}--\eqref{eq:match:7} are gauge invariant.
By computing the matching in Fermi gauge,
we could demonstrate an exact cancellation of the gauge fixing parameters
(cf.\ similarly~\cite{Croon:2020cgk,Schicho:2021gca,Hirvonen:2021zej}).

%
\section{Renormalisation group improvement}
\label{sec:RG-improvement}

At high temperature,
the running with respect to the 4d RG scale $\LamD$,
described by beta-functions $\beta(g^2)$, is not altered compared to zero temperature.
However, an overall RG improvement for the effective potential at high temperature is 
more subtle because
thermal scale hierarchies reflect to the structure of
the potential~\cite{Farakos:1994kx,Gould:2021oba}.
In the mapping between 4d and 3d EFT (cf.\ sec.~\ref{sec:DR}),
the 4d RG scale $\LamD$ cancels between
LO running and
explicit logarithms within the $L_b$-terms at NLO in
the matching relations of the 3d parameters. 
This corresponds to cancellation of the RG scale related to hard mode contributions in 
eq.~\eqref{eq:VThermalCouplingExpansion}.

In the 3d EFT perturbation theory constitutes another, independent,
renormalisation scale $\Lamd$.
Without higher dimensional operators that appear at $\mathcal{O}(g^6)$,
the 3d EFT is super-renormalisable.
Hence,
counterterms in dimensional regularisation have a finite number of terms
sufficient to cancel all UV divergences at any loop order.
In the SM -- and many of its extensions --
only the mass parameters (or tadpole parameters) require renormalisation.
The 3d couplings are RG invariant and
the mass counterterms are exact already at two-loop order.
This gives rise to the exact running of mass parameters in terms of $\Lamd$.
Since the running of the mass appears at two-loop order,
the cancellation of $\Lamd$ in eq.~\eqref{eq:VThermalCouplingExpansion}
leap frogs over even and odd terms.
The running of $g^2$-terms cancels
logarithms at $g^4$-terms, and 
the running of $g^3$-terms cancels
logarithms at $g^5$-terms and so forth. 

In this context,
the discussion in ref.~\cite{Chiang:2017nmu} on
the RG improvement of the cxSM is incomplete.
Therein, the one-loop effective potential is divided into
a tree-level piece,
the zero-temperature Coleman-Weinberg (CW) potential, and
the one-loop thermal function.
Consequently,
the parameters in the tree-level potential are replaced by
the parameters solved from one-loop RGEs
which correctly eliminates
the explicit logarithmic RG scale dependence in the CW potential.
Contrary to~\cite{Chiang:2017nmu}
the renormalization scale does 
enter the high-$T$ effective potential.
While it is true that the one-loop thermal function is
not explicitly dependent on the renormalisation scale,
there is still an implicit running of the parameters inside the thermal function.
The latter, contributes at same order as running of tree-level potential.
Crucially, should one implement
high temperature expansion to this thermal function,
it is straightforward to observe that its leading behaviour
of the quadratic terms -- that contribute to one-loop thermal masses --
is of same order in formal power counting as tree-level terms,
i.e.~$\mathcal{O}(g^2)$.
The running of the one-loop thermal mass contributions is therefore
an effect of $\mathcal{O}(g^4)$ and this running is
cancelled by logarithmic terms for
thermal masses that appear at {\em two-loop} order.
It is indeed these contributions that we include in our analysis with
a NLO dimensional reduction.
We also demonstrated their crucial importance in our numerical analysis of
sec.~\ref{sec:numerics}.

%
\renewcommand{\thesection}{Epilogue}
\tocless\section{Standard Model with a light Higgs}
\label{sec:light-higgs}
\begin{figure}[t]
  \centering
  \includegraphics[width=0.5\textwidth]{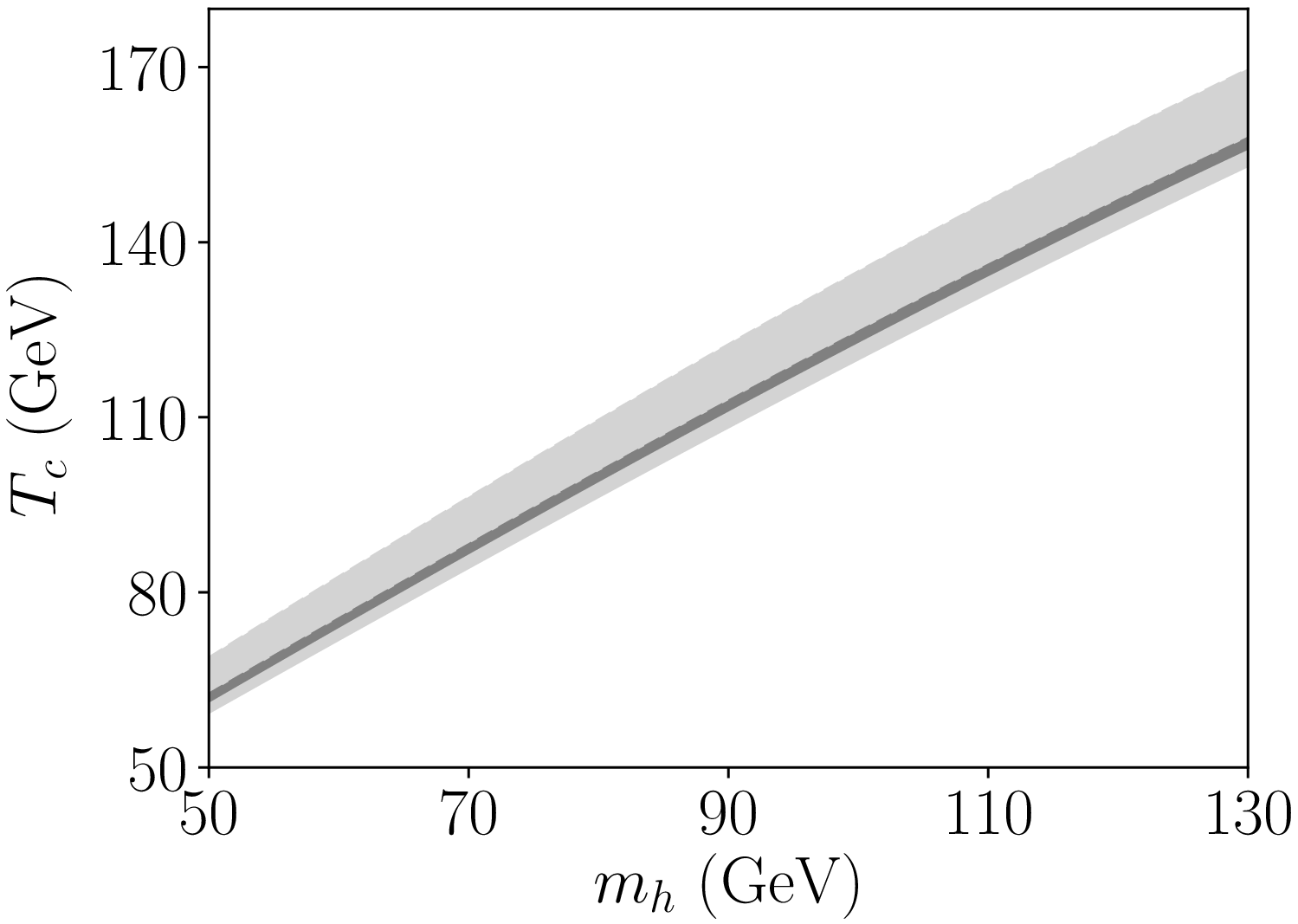}%
  \includegraphics[width=0.5\textwidth]{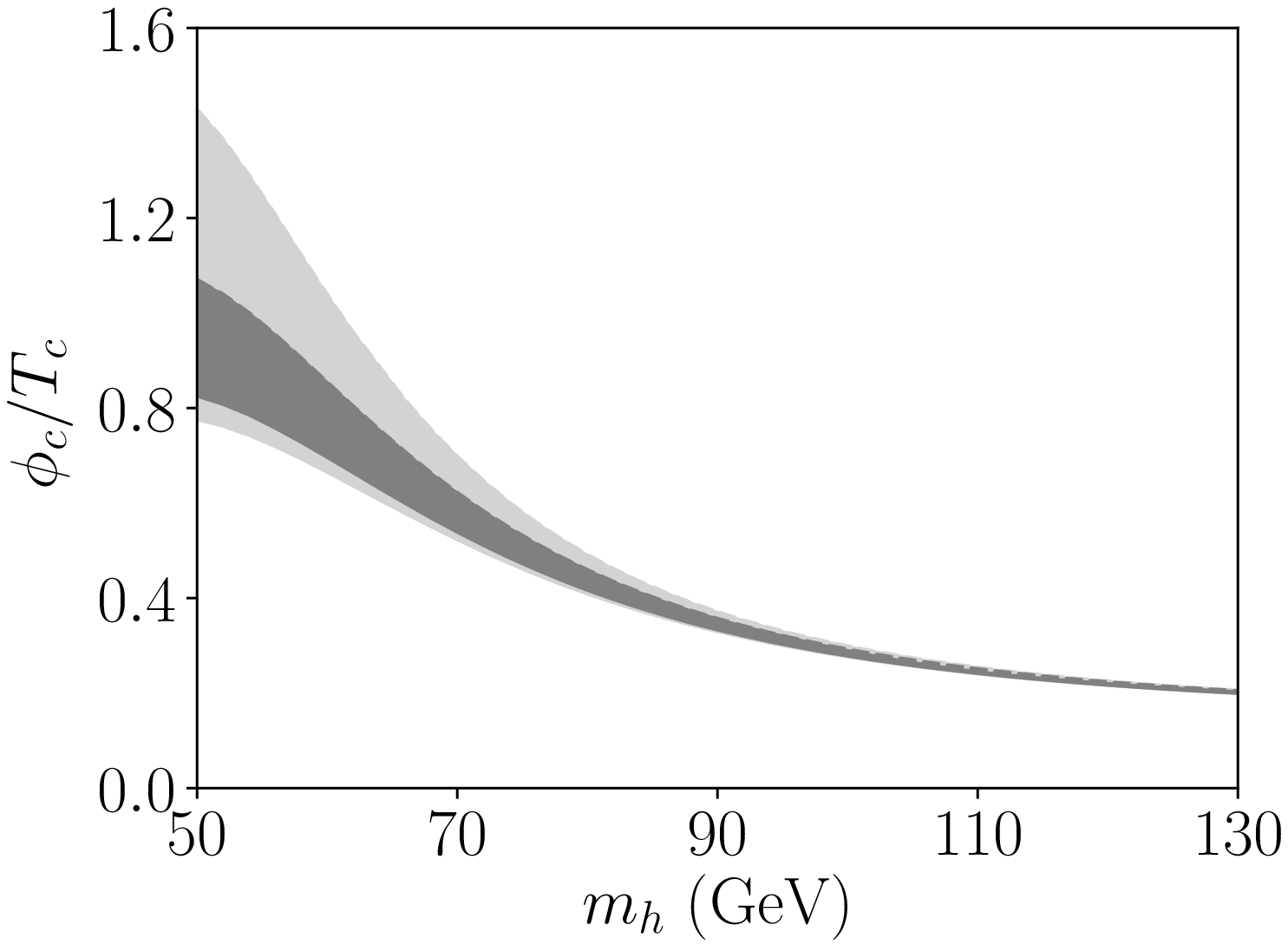} 
  \caption{%
    Left:
    $\Tc$ in a toy SM as a function of a light Higgs mass, in Landau gauge.
    Bands illustrate intrinsic uncertainty related to varying 4d RG scale in
    $\LamD = (0.5 \dots 2.0)\times \pi T$,
    with
    two- (dark grey) and
    one-loop (light grey) thermal mass. 
    Right: As left panel but for $\phi_{\rmi{c}}/\Tc$. 
}
\label{fig:toy-SM}
\end{figure}

\noindent
As an illuminating example, we discuss how even in the pure SM
the running of the one-loop thermal mass can cause
an alarming leftover scale dependence if
the full two-loop thermal mass is not computed. 
For illustration, we work with a toy model of a pure SM
(i.e.\ we omit complex singlet contributions from
the dimensional reduction and 3d EFT of earlier sections) and
we vary the Higgs mass in $m_h = (50\dots 130)$~GeV.  
For simplicity, we determine
the critical temperature $\Tc$ from the condition that
the minima of the one-loop effective potential in 3d EFT are degenerate.
We use the ratio $\phi_{\rmi{c}}/\Tc$ as an estimate of
the transition strength. 
By doing so, we provide an analogy to many 
BSM studies 
that analyse the electroweak phase transition simply in Landau gauge in terms of
the gauge-dependent $\Tc$ and $\phi_{\rmi{c}}/\Tc$. 
In fig.~\ref{fig:toy-SM} we depict these quantities as a function of the Higgs mass. 

The leftover 4d scale dependence is illustrated by the resulting bands from 
varying $\LamD = (0.5 \dots 2.0)\times \pi T$
and by showing the result with
the two-loop (one-loop) thermal mass in dark grey (light grey).
A larger band in the latter case
signals a larger intrinsic uncertainty, and
reinforces our key message of the importance of the two-loop thermal mass.
In BSM theories that involve large portal couplings to
the Higgs, this theoretical uncertainty related to
a leftover scale dependency can be even worse.
In the SM,
the top quark contributions dominate over those related to gauge couplings and
the Higgs self-interaction.
We note, that
the running of the top quark contribution in
the one-loop thermal mass that causes the dominant effect on the broader band,
is exactly the same contribution that causes a similar uncertainty
for a SMEFT study~\cite{Croon:2020cgk},
since this model has exactly the same one-loop thermal mass.
The dominant uncertainty in SMEFT roots in the SM sector and
is not related to the new higher dimensional sextic Higgs operator.

From fig.~\ref{fig:toy-SM}~(right), we observe that
the transition strength increases for lower Higgs mass
which corresponds to smaller Higgs self-coupling and 
a smaller dimensionless 3d quantity
$x \equiv \lambda_{h,3}^{ }/g^2_3$.
This quantity $x$ is the expansion parameter of
3d perturbation theory~\cite{Farakos:1994kx,Farakos:1994xh}.
The smaller it is, the better 3d perturbation theory works.%
\footnote{
  A similar study for a simpler setup of a real scalar theory~\cite{Gould:2021dzl}
  found the explicit 3d expansion parameter and
  inspected convergence and RG improvement within 3d perturbation theory up to three-loop order.
}
However, for smaller $x$
the 3d matching relations are also relatively more sensitive to
the 4d renormalisation scale, and in the end
the total uncertainty is larger.
This provides a counter-example for the common folklore
which states that strong phase transitions (with large $\phi_{\rmi{c}}/\Tc$) are
better described in perturbation theory, than weak transitions.
On the other hand, fig.~\ref{fig:toy-SM} makes it evident that perturbation theory is
oblivious to the crossover character of the SM phase transition after an
end-point around $m_h \sim 70$~GeV~\cite{Kajantie:1995kf}.
There is still a discontinuity in critical value $\phi_{\rmi{c}}$
(indicated by its non-zero value).
In fact,
perturbation theory predicts a first-order phase transition and fails
even at a qualitative level.
This showcases the theoretical challenge to describe phase transition thermodynamics.
   
%
{\small
\bibliographystyle{utphys}
\bibliography{mm}
}
\end{document}